\documentclass[iop,revtex4,appendixfloats]{emulateapj}

\usepackage{graphicx, rotate}
\usepackage{natbib, lscape}
\usepackage{latexsym, amssymb, verbatim}
\usepackage{eufrak}
\usepackage{amsmath}
\usepackage{epsfig}
\usepackage{times}
\usepackage{rotating}
\usepackage{graphicx}
\usepackage{color}
\usepackage{multirow}

\received{ -- }
\revised{ -- }
\accepted{ -- }

\shorttitle{CR-induced destruction of CO in SF galaxies}
\shortauthors{Bisbas et al.}

\newcounter{chem}
\newcounter{temp}

\newcommand{\pasa}{PASA}

\begin{document}

\title{Cosmic-ray induced destruction of CO in star-forming galaxies}


\author{Thomas G. Bisbas\altaffilmark{1,2}, Ewine F. van Dishoeck\altaffilmark{1,3}, Padelis P. Papadopoulos\altaffilmark{4,5,6,7}, L\'aszl\'o Sz{\H u}cs\altaffilmark{1}, Shmuel Bialy\altaffilmark{8} and Zhi-Yu Zhang\altaffilmark{7,9}}

\altaffiltext{1}{Max-Planck-Institut f\"ur Extraterrestrische Physik, Giessenbachstrasse 1, D-85748 Garching, Germany}
\altaffiltext{2}{Department of Astronomy, University of Florida, Gainesville, FL 32611, USA}
\altaffiltext{3}{Leiden Observatory, Leiden University, PO Box 9513, NL-2300 RA Leiden, the Netherlands}
\altaffiltext{4}{School of Physics and Astronomy, Cardiff University, Queen's Buildings, The Parade, Cardiff, CF24 3AA, UK}
\altaffiltext{5}{Research Center for Astronomy, Academy of Athens, Soranou Efesiou 4, GR-115 27 Athens, Greece}
\altaffiltext{6}{Department of Physics, Section of Astrophysics, Astronomy and Mechanics, Aristotle University of Thessaloniki, Thessaloniki 54124, Greece}
\altaffiltext{7}{European Southern Observatory, Headquarters, Karl-Schwarzschild-Strasse 2, D-85748, Garching bei M\"unchen, Germany}
\altaffiltext{8}{Raymond and Beverly Sackler School of Physics \& Astronomy, Tel Aviv University, Ramat Aviv, 69978, Israel}
\altaffiltext{9}{Institute for Astronomy, University of Edinburgh, Royal Observatory, Edinburgh, EH9 3HJ, UK}
\email{TGB: tbisbas@ufl.edu}

\begin{abstract}

We explore the effects of the expected higher cosmic ray (CR) ionization rates $\zeta_{\rm CR}$ on the abundances of carbon monoxide (CO), atomic carbon (C), and ionized carbon (C$^+$) in the H$_2$ clouds of star-forming galaxies. The study of \citet{Bisb15} is expanded by:  a) using realistic inhomogeneous Giant Molecular Cloud (GMC) structures, b) a detailed chemical analysis behind the CR-induced destruction of CO, and c) exploring the thermal state of CR-irradiated molecular gas. CRs permeating the interstellar medium with $\zeta_{\rm CR}$$\ga 10\times$(Galactic) are found to significantly reduce the [CO]/[H$_2$] abundance ratios throughout the mass of a GMC. CO rotational line imaging will then show much clumpier structures than the actual ones. For $\zeta_{\rm CR}$$\ga 100\times$(Galactic) this bias becomes severe, limiting the utility of CO lines for recovering structural and dynamical characteristics of H$_2$-rich galaxies throughout the Universe, \emph{including many of the so-called Main Sequence (MS) galaxies} where the bulk of cosmic star formation occurs. Both C$^+$ and C abundances increase with rising $\zeta_{\rm CR}$, with C remaining the most abundant of the two throughout H$_2$ clouds, when $\zeta_{\rm CR}\sim (1-100)\times$(Galactic). C$^+$ starts to dominate for $\zeta_{\rm CR}$$\ga 10^3\times$(Galactic). The thermal state of the gas in the inner and denser regions of GMCs is invariant with $T_{\rm gas}\sim 10\,{\rm K}$ for $\zeta_{\rm CR}\sim (1-10)\times$(Galactic). For $\zeta_{\rm CR}$$\sim 10^3\times$(Galactic) this is no longer the case and $T_{\rm gas}\sim 30-50\,{\rm K}$ are reached. Finally we identify OH as the key species whose $T_{\rm gas}-$sensitive abundance could mitigate the destruction of CO at high temperatures.

\end{abstract}


\section{Introduction}%

Molecular hydrogen (H$_2$) gas and its mass distribution in galaxies is of fundamental importance in determining their structural and dynamical characteristics, as well as the process of star formation in them. It does not have a permanent dipole moment, and at its lowest energy level ($\sim510\,{\rm K}$), the S(2-0) quadrupole transition in the far-IR wavelength cannot trace the bulk of the H$_2$ molecules which predominantly lie in the cold ($\lesssim100\,{\rm K}$) phase. The astronomical community has therefore implemented other lines to infer this mass indirectly typically using CO, the next most abundant molecule after H$_2$ itself with its bright rotational transitions in the millimeter/sub-millimeter wavelength regime \citep[{[}CO{]}/{[}H$_2${]}$\sim 10^{-4}$ in Milky Way, e.g.][ where {[]} denotes the abundance compared to H-nuclei number density]{Lacy94}. Unlike H$_2$, CO has a permanent dipole moment and rotational transitions with $\Delta J=1$ are allowed, e.g. CO $J=1-0$ at $115\,{\rm GHz}$ is the most commonly used as an H$_2$ gas tracer, with higher-$J$ transitions becoming accessible at high redshifts in the age of Atacama Large Millimeter/submillimeter Array (ALMA) at the high altitude of Llano de Chajnantor plateau in Chile. The goal of this work is to explore to what extent CO remains a good tracer of the molecular gas mass and dynamics in regions with elevated CRs, such as expected in actively star-forming galaxies typical for the early Universe.

Once the CO ($J=1-0$) line emission is detected, a scaling factor is used to convert its velocity-integrated brightness temperature (or the line luminosity) to H$_2$ column density on scales of molecular clouds or larger. This method is statistically robust for M(H$_2$)$\ga $10$^5\,{\rm M}_{\sun}$ \citep[for an investigation on the physical condition dependencies and the underlying physics of the CO-to-H$_2$ conversion factor, see e.g.][]{Bola13,Szuc16}. This CO-to-H$_2$ method, calibrated in Galactic conditions \citep{Dick86,Solo87}, is widely used in extragalactic observations \citep[e.g.][]{Solo97,Grat16,Chen15,Genz15}. If multi-$J$ CO (or other molecules like HCN) line observations reveal average gas densities, temperatures and/or dynamic states of molecular clouds that differ from those in the Milky Way there exists a theoretical framework to use appropriately modified CO-to-H$_2$ conversion factors \citep[e.g.][]{Brya96,Papa12a}. All these techniques work as long as CO and other molecules used to study its average conditions (e.g. HCN) remain sufficiently abundant in GMCs, typically not much less abundant  as in the Galactic GMCs where these techniques have been calibrated. Low-metallicity (Z) molecular gas, especially when irradiated by strong FUV radiation, was the  first H$_2$ gas phase for which early studies showed that the standard techniques actually fail \citep{Pak98,Bola99}. This means that low-Z gas in the outer parts of even ordinary spiral galaxies, like the Milky Way, may then be in a very CO-poor phase and thus impossible to trace using CO lines \citep{Papa02,Wolf10}.

Atomic carbon (C) line emission is another alternative for deducing the molecular gas distribution in galaxies and one that can be as reliable as low-$J$ CO lines. This is because of its widespread emission in H$_2$ clouds despite of what is expected from the classical theory of Photodissociation Regions (PDRs) \citep{Geri00,Isra01,Papa04,Bell07,Offn14,Glov15}. There are a  number of reasons contributing towards C line emission being fully associated with CO line emission and having larger emergent flux densities per H$_2$ column density than those of the low-CO rotational lines used as global H$_2$ gas tracers. This led to an early proposal for using the two C lines, $^3$P$_1$-$^3$P$_0$ ($W_{\rm CI,1-0}$) at 492~GHz and $^3$P$_2$-$^3$P$_1$ ($W_{\rm CI,2-1}$) at 809~GHz, and especially the lower frequency line, as routine H$_2$ gas tracers in galaxies for $z\ga 1$ when the lines shift into the millimeter band \citep{Papa04,Papa04b}. Such a method can now be extended in the local Universe as imaging at high frequencies can be performed by ALMA \citep{Krip16}. In our Galaxy, the Vela Molecular Ridge cloud C shows that atomic carbon can trace accurately the H$_2$ gas mass \citep{Lo14}. For extragalactic studies, \citet{Zhan14} find that in the centre of the Seyfert galaxy Circinus, the C-traced H$_2$ mass is consistent with that derived from sub-millimeter dust continuum and multiple-$J$ CO excitation analysis, while C observations have recently been used to trace the H$_2$ gas mass in distant starbursts at z$\sim $4 \citep{Both17}.

The ongoing discussion regarding the widespread C line emission in molecular clouds, and thus their ability to trace H$_2$ independently of $^{12}$CO and $^{13}$CO lines, took another turn after the recent discovery that cosmic rays (CRs) can very effectively destroy CO throughout H$_2$ clouds, leaving C (but not much C$^+$) in their wake \cite[][hereafter `B15', see also \citet{Bial15}]{Bisb15}. Unlike FUV photons that only do so at the surface of H$_2$ clouds and produce C$^+$ rather than C, CRs destroy CO volumetrically and can render H$_2$ clouds partly or wholly CO-invisible even in ISM environments with modestly boosted CR ionization rates of $\zeta_{\rm CR}\sim(10-50)\times$Galactic, where $\zeta_{\rm CR}$ is the cosmic-ray ionization rate (${\rm s}^{-1}$) \citep{Stro04a,Stro04b}. The latter values are expected in typical star-forming (SF) galaxies in the Universe \citep{Hopk06,Dadd10}, currently studied only using CO \citep[e.g.][]{Genz12}. For example, \citet{Mash13} inferred a cosmic-ray ionization rate of $\zeta_{\rm CR}\sim3\times10^{-14}\,{\rm s}^{-1}$ in their analysis of CO/C$^+$ emissions in the high redshift HDF 850.1. B15 found that besides the ability of C lines in tracing the CO-rich parts of an H$_2$ cloud, they also probe the CO-poor regions. This is of particular interest especially if its lines are to be a viable H$_2$-tracing alternative to CO lines. In the current work we re-examine these CR-induced effects discussed by B15 in the much more realistic setting of inhomogeneous H$_2$ clouds, that could affect their `visibility' in CO, C, and C$^+$ line  emission. Furthermore, we discuss in more detail the chemistry behind the CR-control of the [CO]/[H$_2$] abundance ratio and its dependence on the gas temperature which itself is affected by cosmic rays. The latter proves to be a very important factor that should be taken into account in turbulent-dynamic cloud simulations that explore similar issues. 

Models of CO-destruction in cosmic-ray dominated regions (CRDRs), predict that low-$J$ CO/C line flux ratios are general low, $<1$. Recent ALMA observations of the Spiderweb galaxy by \citet{Gull16} find that $W_{\rm CO(7-6)}$/$W_{\rm CI,2-1}\sim\!0.2$ which can be potentially explained from the presence of high CR energy densities. Another interesting recent example is the observation of the $W_{\rm CO(1-0)}$/$W_{\rm CI,1-0}\sim\!0.1-0.4$ ratio in the starburst galaxy NGC253 \citep{Krip16} which, in association with early $W_{\rm CO(7-6)}$ observations indicating warm H$_2$ gas \citep{Brad03}, could be due to high $\zeta_{\rm CR}$ values. High CR energy densities are  expected to maintain higher gas temperatures even in far-UV-shielded environments. B15 estimate a gas temperature of $\sim\!50\,{\rm K}$ when the CR ionization rate, $\zeta_{_{\rm CR}}$, is boosted up to $\sim10^3$ times the mean Galactic value.  

In this paper we perform astrochemical simulations of the effects of larger than Galactic CR energy densities on inhomogeneous molecular clouds, using the {\sc 3d-pdr} code \citep{Bisb12} to infer the distributions of the CO, C, C$^{+}$ abundances and of the gas temperature. This is a continuation of the B15 work using much more realistic molecular cloud structures  rather than those of uniform-density or radially varying densities explored previously. Moreover, we now also analyze the chemistry involved in the CR-induced destruction of CO, and its conversion to C, in greater detail. In all of our simulations we assume that the bulk of the H$_2$ gas interacts with CRs throughout the cloud volume (i.e. the H$_2$ gas `sees' CRs, with the same spectrum, throughout the volume of the cloud). While this is not true for some regions deep inside clouds \citep{Rimm12}, and can depend on the specifics of magnetic fields \citep{Pado13}, it remains a very good approximation for the {\it bulk} of H$_2$ clouds in SF-galaxies \citep{Papa11}.

The paper is organized as follows. In Section~\ref{sec:simulations} we present the setup of our simulations. In Section~\ref{sec:results} we present the results of our calculations and in particular how the probability density functions and the abundance distribution of the above key species, as well as the corresponding heating and cooling functions vary under the different conditions explored. In Section~\ref{sec:oh} we discuss how OH enhances the [CO]/[H$_2$] abundance ratio at higher temperatures when $\zeta_{_{\rm CR}}$ increases and in Section~\ref{sec:discussion} we refer to the impact of our findings in observations. We conclude in Section~\ref{sec:conclusions}.

\section{Description of simulations}%
\label{sec:simulations}
We  consider a three-dimensional density distribution of a non-uniform giant molecular cloud (GMC) and use the {\sc 3d-pdr} \citep{Bisb12} code to perform chemistry and full thermal balance calculations and estimate the abundance distribution of chemical species and the gas temperature distribution.

\subsection{Density distribution}
The inhomogeneous spherical GMC in our models is rendered by a fractal structure with a fractal dimension of ${\cal D}=2.4$ constructed using the method described in \citet{Walc15}. It has a radius of $R=10\,{\rm pc}$ and mass of $M=1.1\times10^5\,{\rm M}_{\odot}$. This corresponds to an average H-nucleus number density of $\langle n\rangle\simeq760\,{\rm cm}^{-3}$, typical for Milky Way GMCs. The central part of the cloud contains a dense region with peak density $\sim2\times10^4\,{\rm cm}^{-3}$. The fractal dimension is in accordance to the clumpiness factor observed in evolved Galactic H{\sc ii} regions \citep[e.g.][]{Sanc10, Walc15}. On the contrary, for diffuse clouds the fractal dimension is higher (${\cal D}\sim2.8-3.0$) meaning that they are more uniform \citep{Walc15}. The chosen dimension of ${\cal D}=2.4$ corresponds to a GMC containing non-homogeneously distributed high density clumps typical of those that eventually undergo star formation. They are therefore expected to be H$_2$-rich and for the particular Milky Way conditions, also CO-rich. We do not evolve the cloud hydrodynamicaly and in order to resolve its densest parts, we use a Smoothed Particle Hydrodynamics setup of the cloud and represent it with $8.33\times10^5$ particles\footnote{The density of each particle is calculated using the SPH code {\sc seren} \citep{Hubb11}.}.

\begin{figure*}
\center
\includegraphics[width=0.985\textwidth]{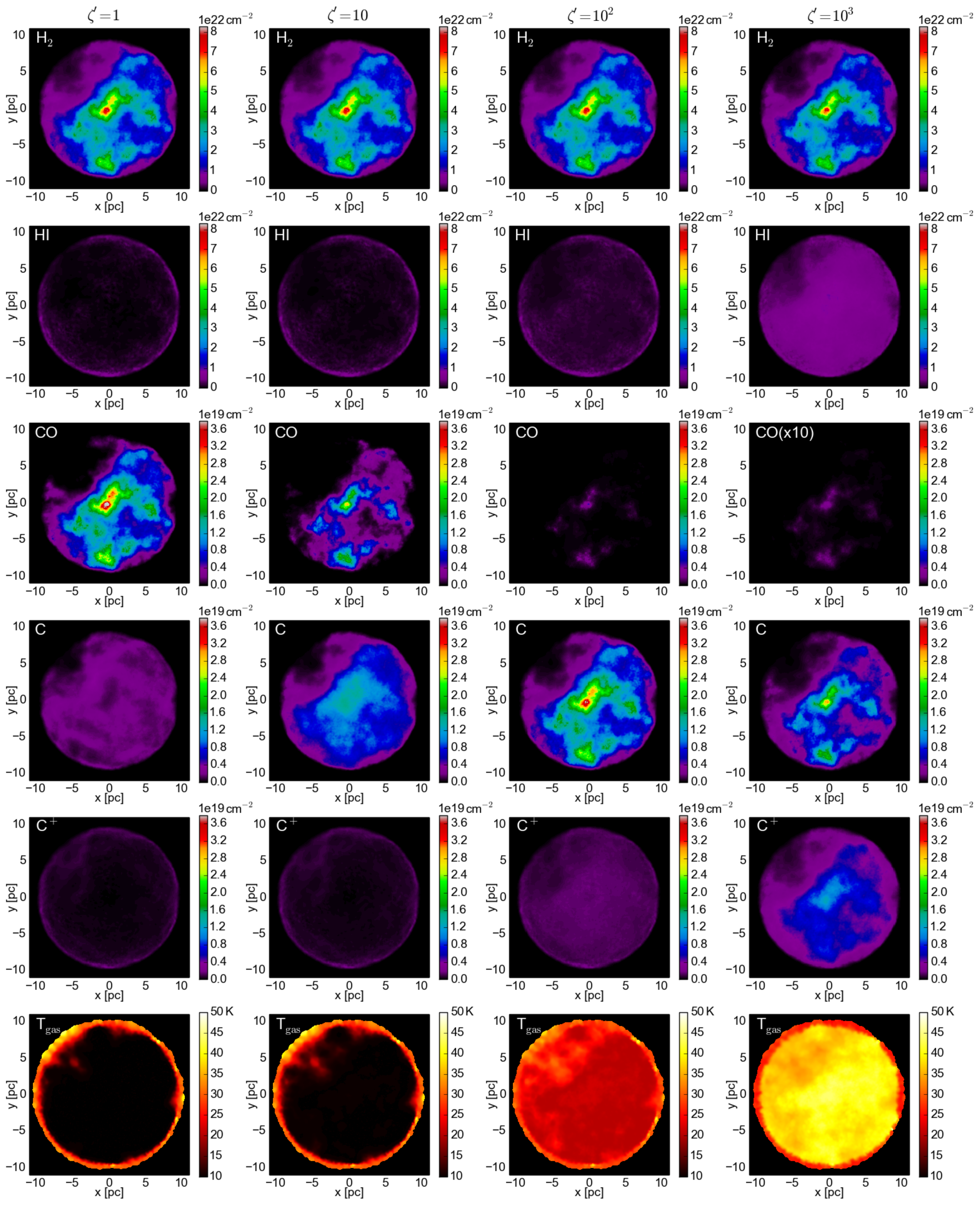}
\caption{ Column density ($N$) plots of H$_2$ (top row), H{\sc i} (second row), CO (third row), C (fourth row) and C$^+$ (fifth row). The colour bar has units of ${\rm cm}^{-2}$ and the axes have units of ${\rm pc}$. From left to right, $\zeta'=1,10,10^2,10^3$. 
Note that $N$(CO) at $\zeta'=10^3$ is raised 10 times to make the structure visible. 
As $\zeta'$ increases, $N$(H$_2$) remains remarkably similar whereas $N$(CO) is reduced by approximately one order of magnitude. At the same time $N$(C) peaks for $\zeta'=10^2$ while for $\zeta'>10^2$ it is transformed into C$^+$. It is interesting to note that $N$(C) at $\zeta'=10^2$ is approximately equivalent to $N$(CO) at $\zeta'=1$. The bottom row shows cross sections of the gas temperature at the $z=0\,{\rm pc}$ plane. The colour bar there has units of K. The gas temperature at the interior of the cloud increases with $\zeta'$, reaching values up to $\simeq50\,{\rm K}$. For $\zeta'=1$, $T_{\rm gas}\simeq10\,{\rm K}$ in the cloud centre as observed in Milky Way. In all cases the external shell is irradiated by the isotropic FUV radiation and thus its temperature is determined by that interaction.}
\label{fig:cd}
\end{figure*}

\begin{figure}
\center
\includegraphics[width=0.23\textwidth]{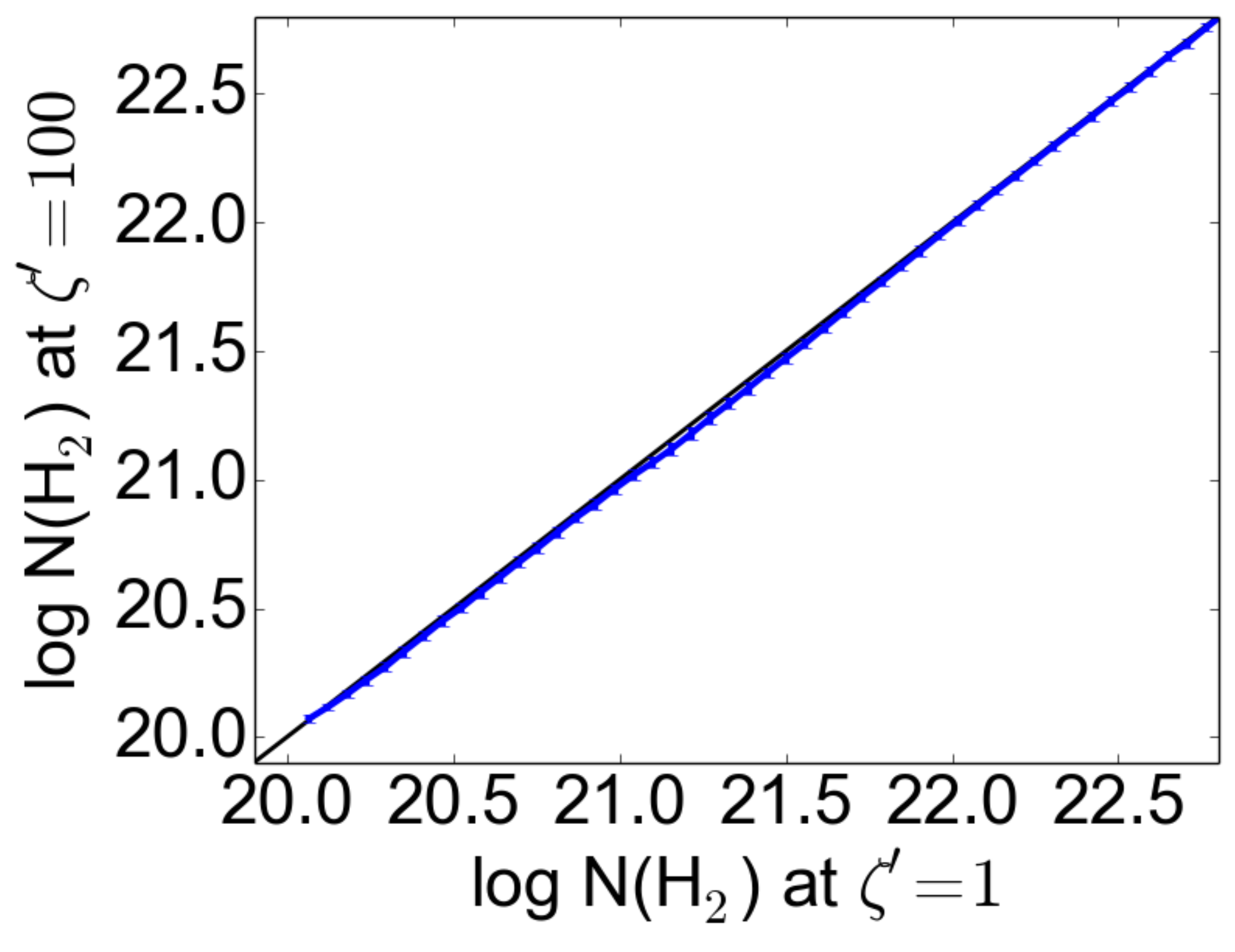}
\includegraphics[width=0.23\textwidth]{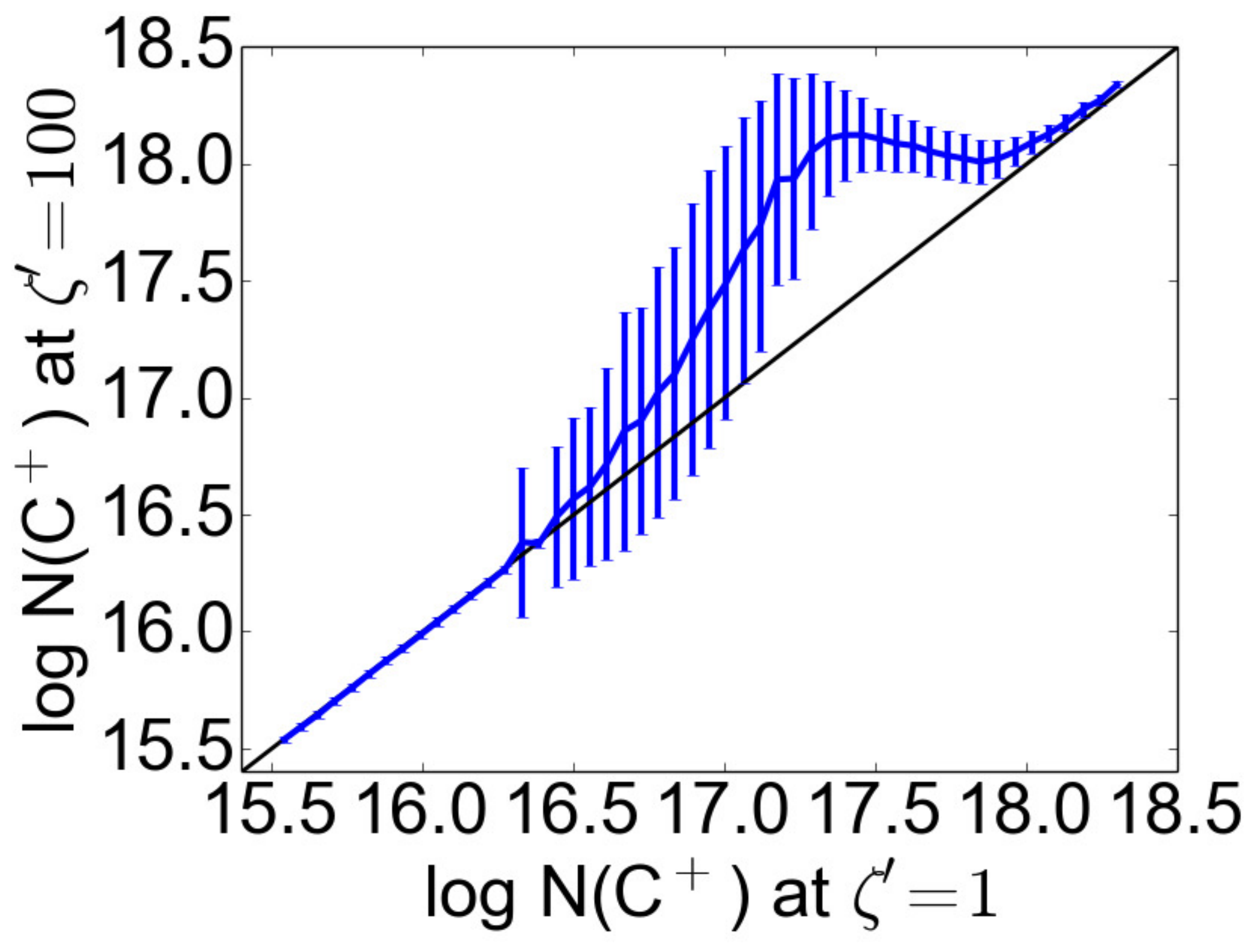}
\includegraphics[width=0.23\textwidth]{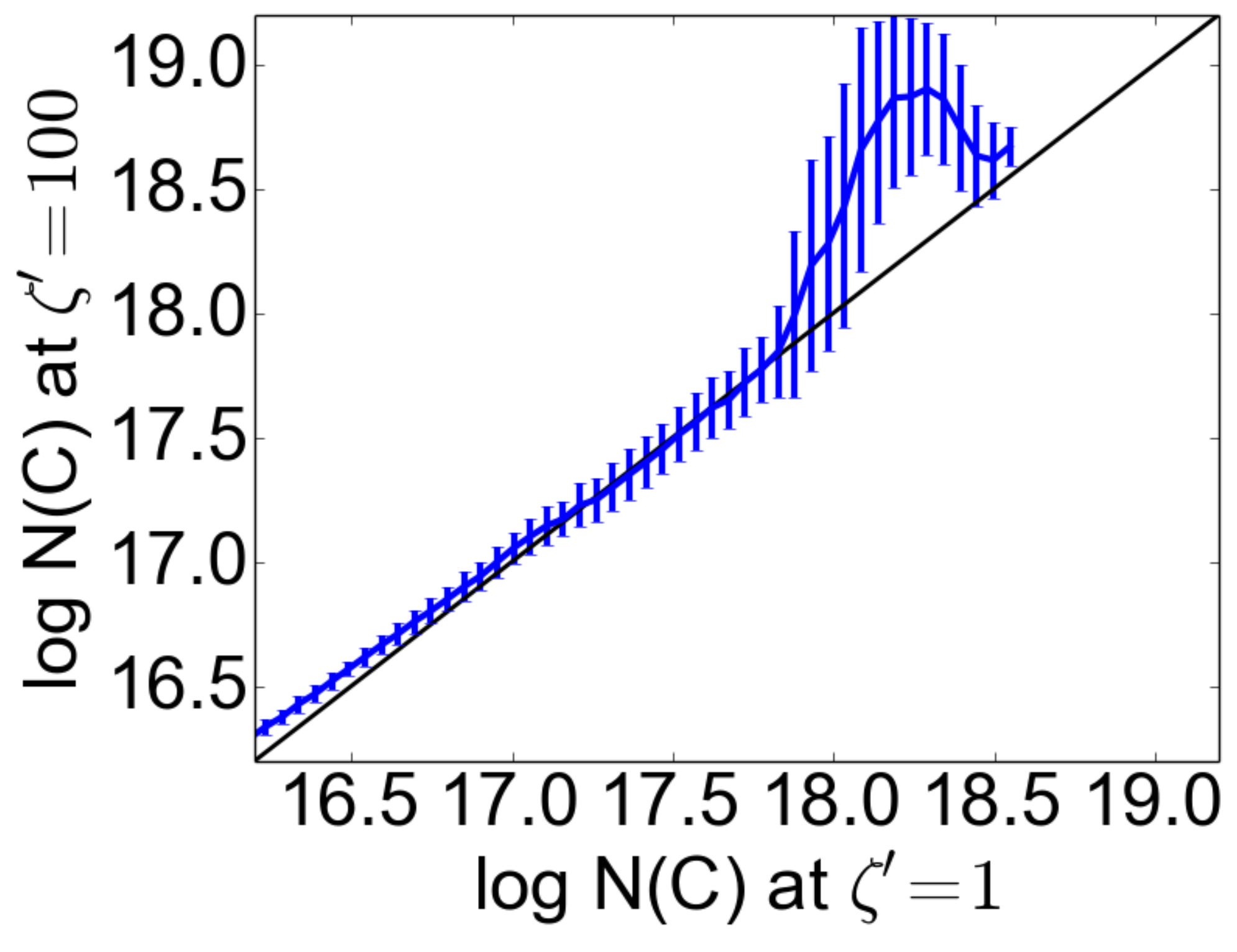}
\includegraphics[width=0.23\textwidth]{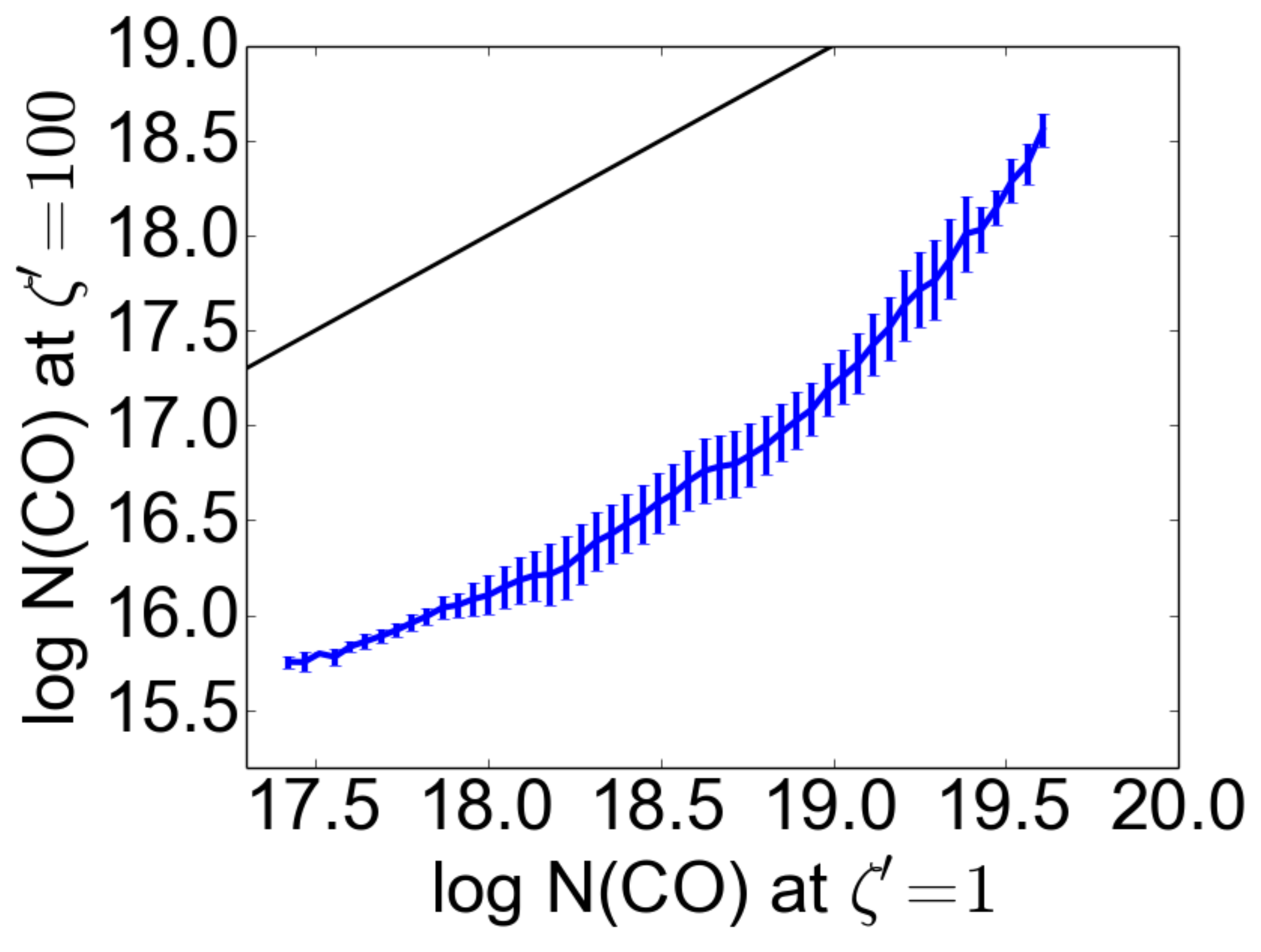}
\caption{ Comparison of column densities at $\zeta'=1$ versus $\zeta'=10^2$ for H$_2$ (top left), C$^+$ (top right), C (bottom left), and CO (bottom right). These four panels illustrate Eqns.~\ref{eqn:rel1}-\ref{eqn:rel2}, extracted from the simulations shown in Fig.~\ref{fig:cd}. The black solid line is the $y=x$ relation to guide the eye. The error bars correspond to $1\sigma$ standard deviation. These plots show that nearly all CO that has been destroyed by CRs is converted to C and C$^+$, while H$_2$ remains remarkably unaffected.}
\label{fig:relations}
\end{figure}

\begin{table}
\caption{ Initial gas-phase chemical abundances used in the present paper. The abundances correspond to Solar undepleted abundances \citep{Aspl09} and are relative to the total hydrogen nuclei number density.}.
\centering
\label{tab:abun}
\begin{tabular}{l l l l}
\hline
\hline
H & $4.00\times10^{-1}$ & Mg$^+$ & $3.98\times10^{-5}$ \\
H$_2$ & $3.00\times10^{-1}$ & C$^+$ & $2.69\times10^{-4}$ \\
He & $8.50\times10^{-2}$ & O & $4.90\times10^{-4}$ \\
S & $1.32\times10^{-5}$ & & \\
\hline
\end{tabular}
\end{table}


\subsection{{\sc 3d-pdr} initial conditions}
\label{ssec:ics}
We use the {\sc 3d-pdr} code \citep{Bisb12} in order to calculate the abundances of chemical species in the above fractal cloud. {\sc 3d-pdr} obtains the gas temperature and the abundance distribution of any arbitrary three-dimensional density distribution by balancing various heating and cooling functions (see \S\ref{ssec:heatcool}). For the simulations of this work we use the same chemical network and initial abundances of species as used in the B15 paper. In particular we use a subset of the UMIST 2012 network \citep{McEl13} consisting of 6 elements (H, He, C, O, Mg, S), 58 species and more than 600 reactions. Table \ref{tab:abun} shows the initial abundances used which correspond to undepleted Solar abundances with hydrogen mostly in molecular form \citep{Aspl09}. We chemically evolve the cloud for $t_{\rm chem}=10^7\,{\rm yr}$ at which point the system has reached chemical equilibrium. Chemical equilibrium is typically obtained after $t_{\rm chem}\sim10^5\,{\rm yr}$ for a cloud in which H$_2$ has already formed, \citep[e.g.,][]{Bell06}, which is comparable to turbulent diffusion timescales for GMCs in ULIRG environments \citep[][see also \S5.1 of B15]{Xie95,Papa04}. For our modelled GMC, we find that the sound crossing time is $\sim3\,{\rm Myr}$. On the other hand the H$_2$ formation time is $t_{\rm chem}=1/Rn_{\rm H}\la5\,{\rm Myr}$, where $R=3\times10^{-18}(T_{\rm gas}/{\rm K})^{1/2}\,{\rm cm}^3\,{\rm s}^{-1}$. We therefore do not expect turbulence to strongly affect our results, although hydrodynamical simulations exploring this effect are needed towards this direction \citep[e.g.,][]{Glov16}. We include H$_2$ formation on dust grains but we do not model CO freeze-out. The effects of different networks and different elemental abundances are presented in Appendix \ref{app:chem}, which shows that our trends are robust.

In all simulations we consider an isotropic FUV radiation field strength of $\chi/\chi_0=1$, normalized to the \citet{Drai78} spectral shape and width is equivalent to $2.7\times10^{-3}\,{\rm erg}\,{\rm cm}^{-2}\,{\rm s}^{-1}$ integrated over the $91.2-240\,{\rm nm}$ wavelength range \citep{Habi68}. At the surface of the cloud, the field strength is therefore approximately equal to $1/4\,{\rm Draine}$ \citep{Ster14}. We perform a suite of four simulations by varying the cosmic-ray ionization rate, $\zeta_{_{\rm CR}}$ from $10^{-17}\,{\rm s}^{-1}$ to $10^{-14}\,{\rm s}^{-1}$, the upper limit of which corresponds to values suggested for the Central Molecular Zone \citep[e.g.][]{LePe16}. For convenience we normalize $\zeta_{_{\rm CR}}$ as
\begin{eqnarray}
\label{eqn:zeta'}
\zeta'\equiv\zeta_{_{\rm CR}}/\zeta_{_{\rm MW}}, 
\end{eqnarray}
where $\zeta_{_{\rm MW}}=10^{-17}\,{\rm s}^{-1}$ is the typically adopted ionization rate of the Milky Way. This latter value is $\sim0.1$ times that observed in the diffuse ISM \citep[e.g.][]{McCa03,Dalg06,Neuf10,Indr12,Indr15} but close to the Heliospheric value ($1.45-1.58\times10^{-17}\,{\rm s}^{-1}$) as measured by the Voyager 1 spacecraft \citep{Cumm15}. Our baseline choice of a value lower than that observed is made under the assumption that cosmic rays in our model do not attenuate as a function of column density; instead the corresponding ionization rate remains constant everywhere in the cloud. We therefore adopt a baseline value that corresponds to an already attenuated $\zeta_{_{\rm CR}}$ within denser H$_2$ gas\footnote{A similar approximation has also been made by \citet{Nara16}}. 

\begin{figure}
\center
\includegraphics[width=0.45\textwidth]{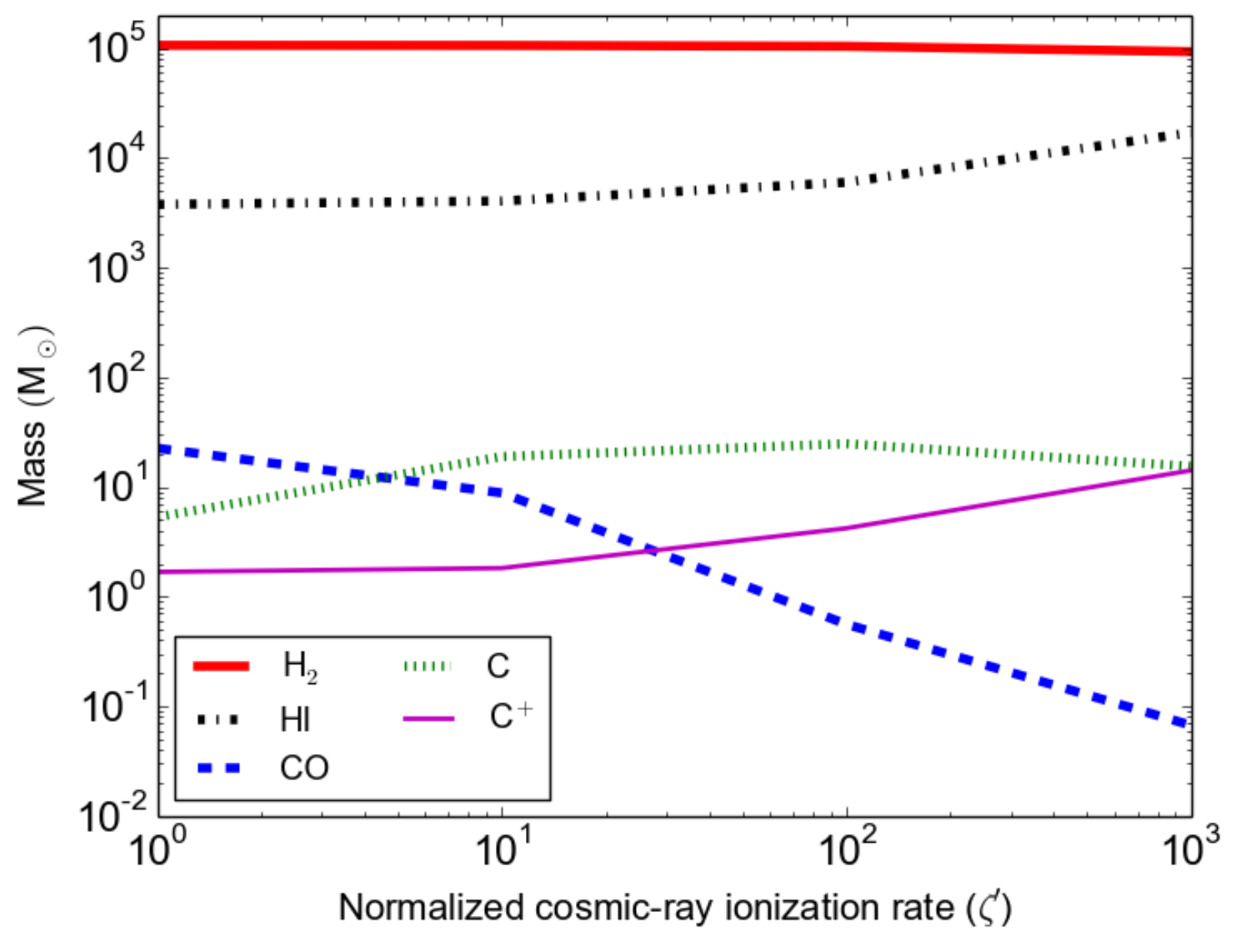}
\caption{ Dependency of M(H$_2$) (red thick solid line), M(H{\sc i}) (black dot-dashed line), M(CO) (blue dashed line), M(C) (green dotted line) and M(C$^+$) (magenta thin solid line) as a function of $\zeta'$. As $\zeta'$ increases, M(H$_2$) monotonically decreases while M(H{\sc i}) and M(C$^+$) monotonically increase. M(C) appears to have a local maximum at $\zeta'\sim10^2$. For $\zeta'\gtrsim10^2$, M(CO) is $\sim2$ orders of magnitude less than that of $\zeta'=1$.}
\label{fig:masses}
\end{figure}

\subsection{Cosmic-ray ionization rate and UV}
High cosmic ray ionization rates, on the order of $\zeta'=10^3$, are expected in starburst environments such as the (ultra-) luminous infrared galaxies \citep[U/LIRGs, i.e. $L_{\rm IR}>10^{11}-10^{12}L_{\odot}$][]{Sand03}. In these systems the star formation rate (SFR) density, $\rho_{_{\rm SFR}}\equiv{\rm SFR}/V$ (where SFR is in $M_{\odot}/{\rm year}$ and $V$ is the corresponding volume), is enhanced by a factor up to $\sim10^3$ compared with the Milky Way. This drives a higher cosmic ray energy density as $U_{_{\rm CR}}\propto\rho_{_{\rm SFR}}$ \citep{Papa10}. Enhanced FUV fields are also expected in such environments, although dust attenuation in these metal-rich objects will keep the boost of the {\it average} FUV field incident on the H$_2$ clouds lower than proportional to $\rho_{_{\rm SFR}}$ \citep{Papa14}.

 In this paper we do not vary the isotropic FUV radiation field in our simulations, wanting to isolate the effects of CRs. We note, however, that chemo-hydrodynamical simulations performed by \citet{Glov16} suggest that if both $\zeta'$ and $\chi$ are increased by two orders of magnitude, clouds with mass $M\sim10^4\,{\rm M}_{\odot}$ might be dispersed by the thermal pressure which would dominate over the gravitational collapse. The attenuation of the FUV radiation is calculated using the method described in \citet{Bisb12} which accounts for the attenuation due to dust,  H$_2$ self-shielding CO self-shielfing, CO shielding by H$_2$ lines and CO shielding by dust.

\begin{figure}
\center
\includegraphics[width=0.42\textwidth]{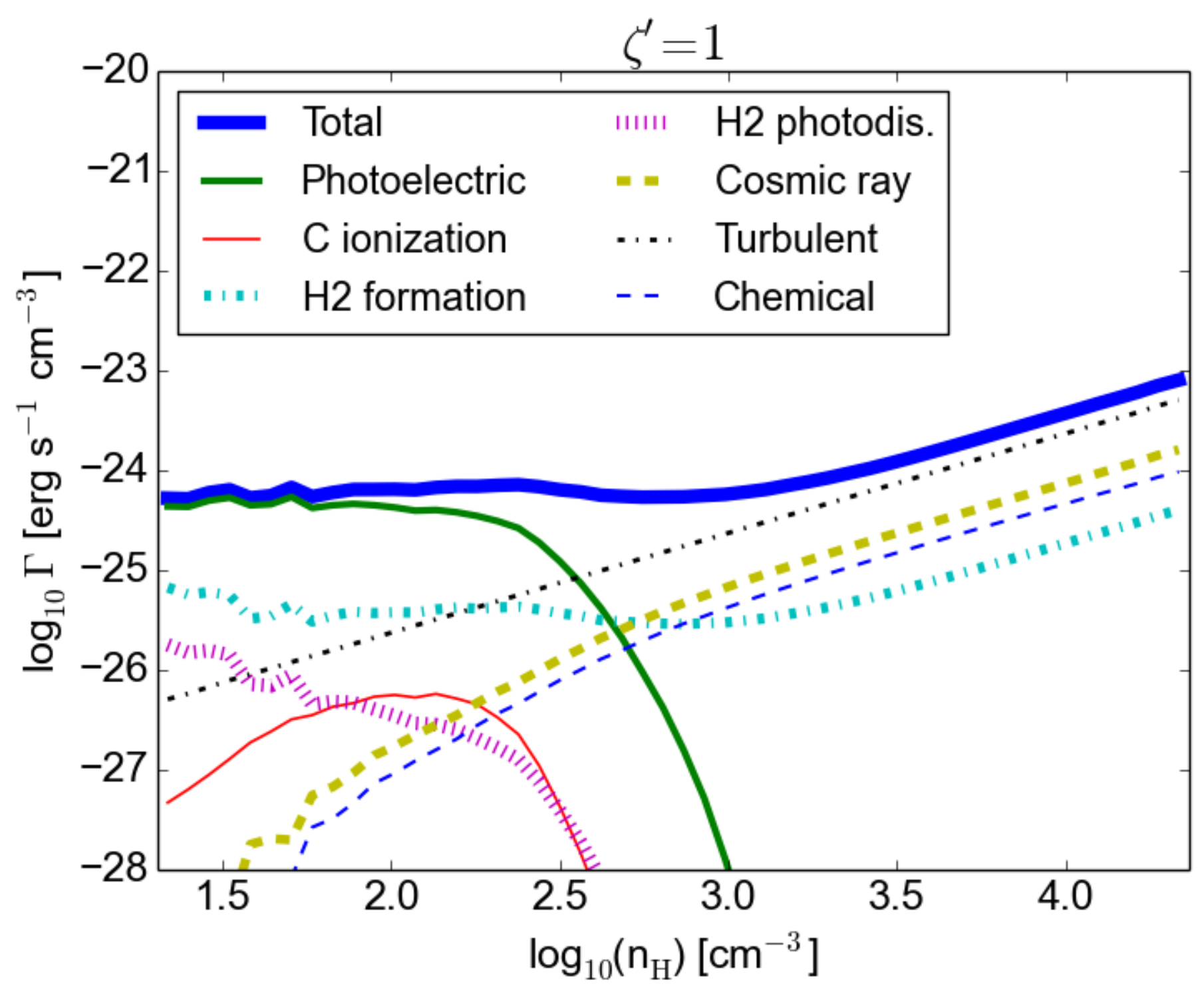}
\includegraphics[width=0.42\textwidth]{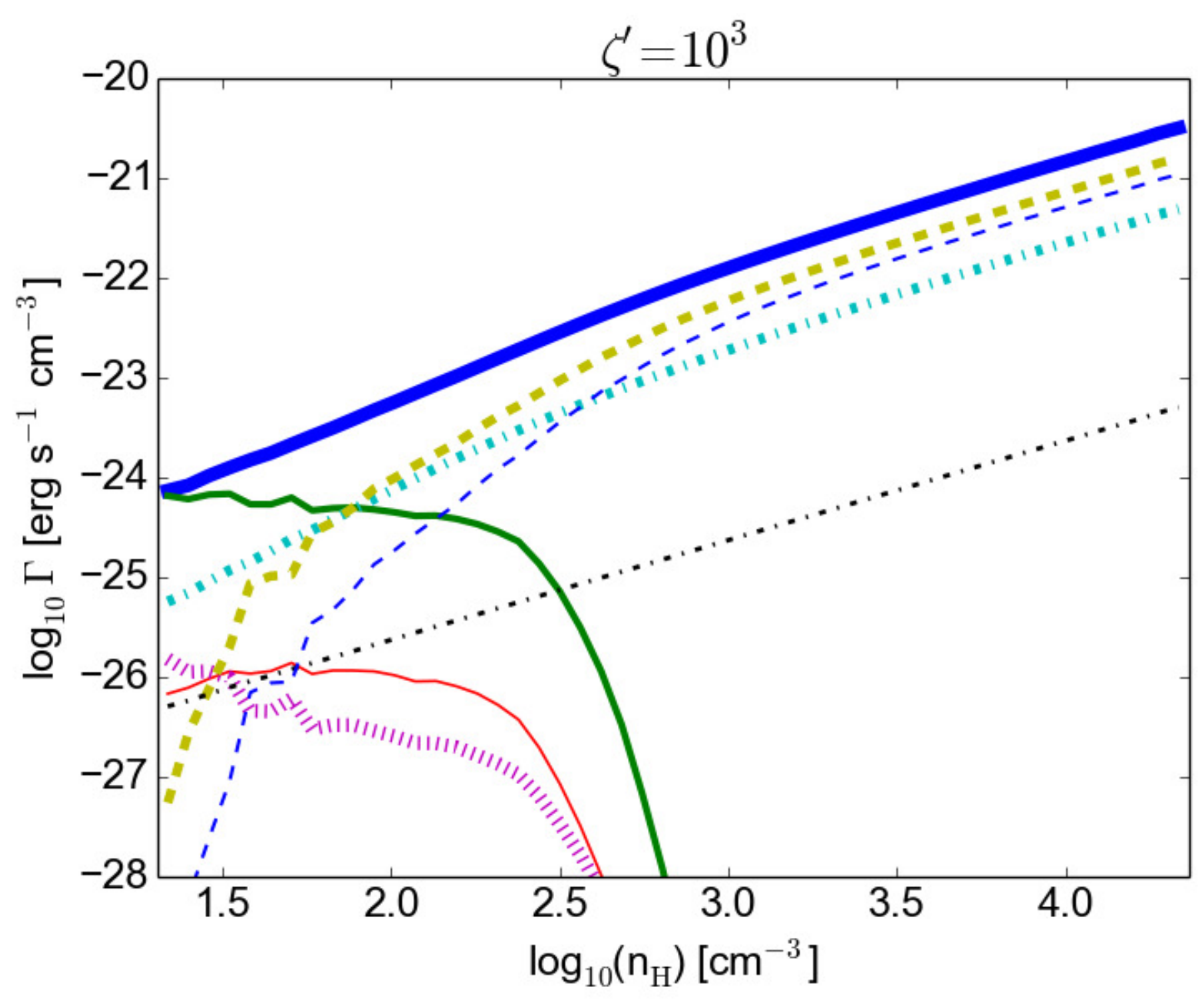}
\caption{ Heating processes for $\zeta'=1,10^3$. The curves correspond to the mean values for the entire density distribution. For low densities which are located at low visual extinction, heating is, in both cases, predominantly a result of the photoelectric effect due to the interaction of the isotropic FUV field and dust grains. For higher densities, at $\zeta'=1$ heating due to turbulent dissipation is the main contributor to the total heating rate while at $\zeta'=10^3$ cosmic ray, chemical, and H$_2$ formation heating are responsible for the increase of $T_{\rm gas}$.}
\label{fig:heat}
\end{figure}

\begin{figure}
\center
\includegraphics[width=0.42\textwidth]{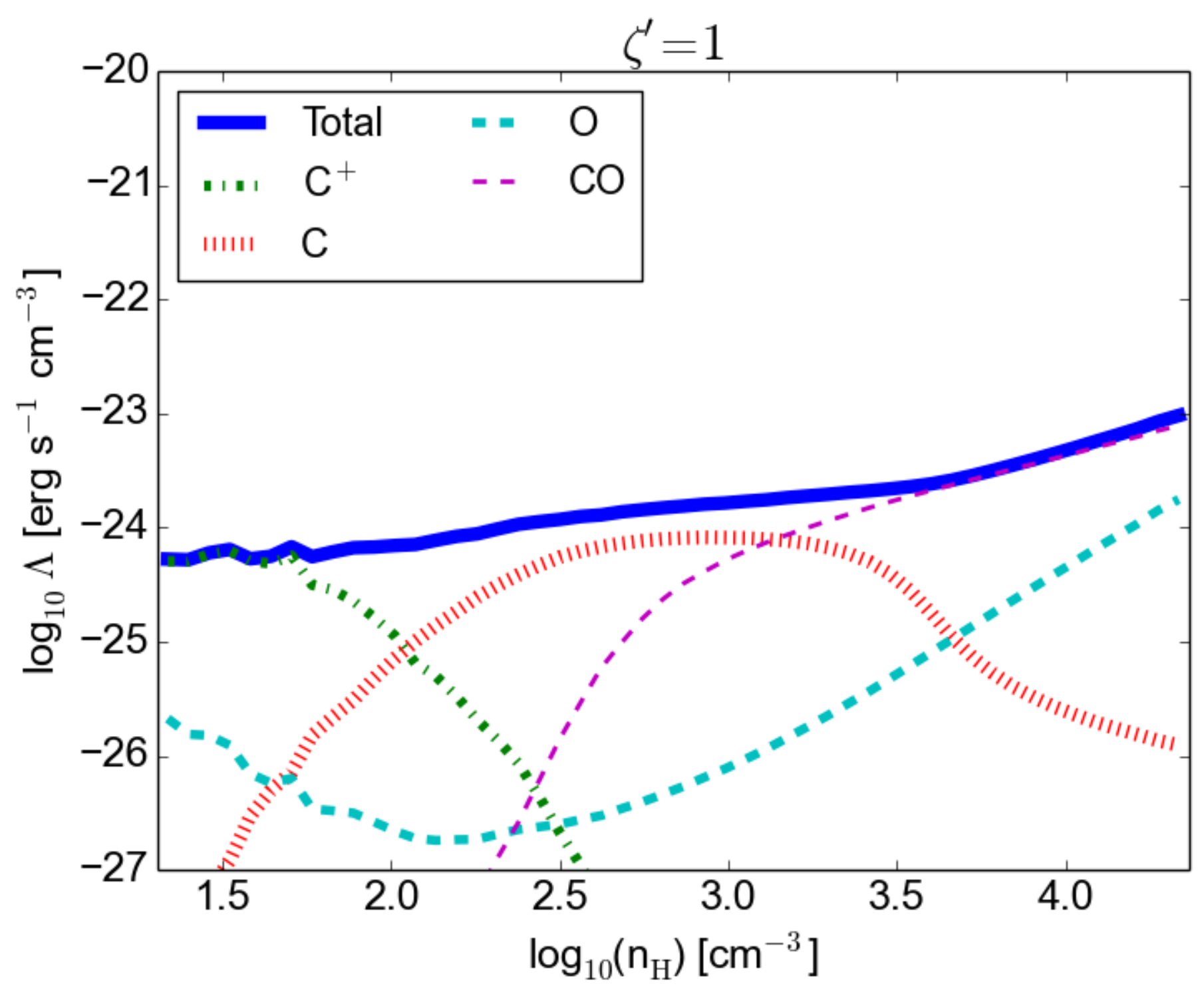}
\includegraphics[width=0.42\textwidth]{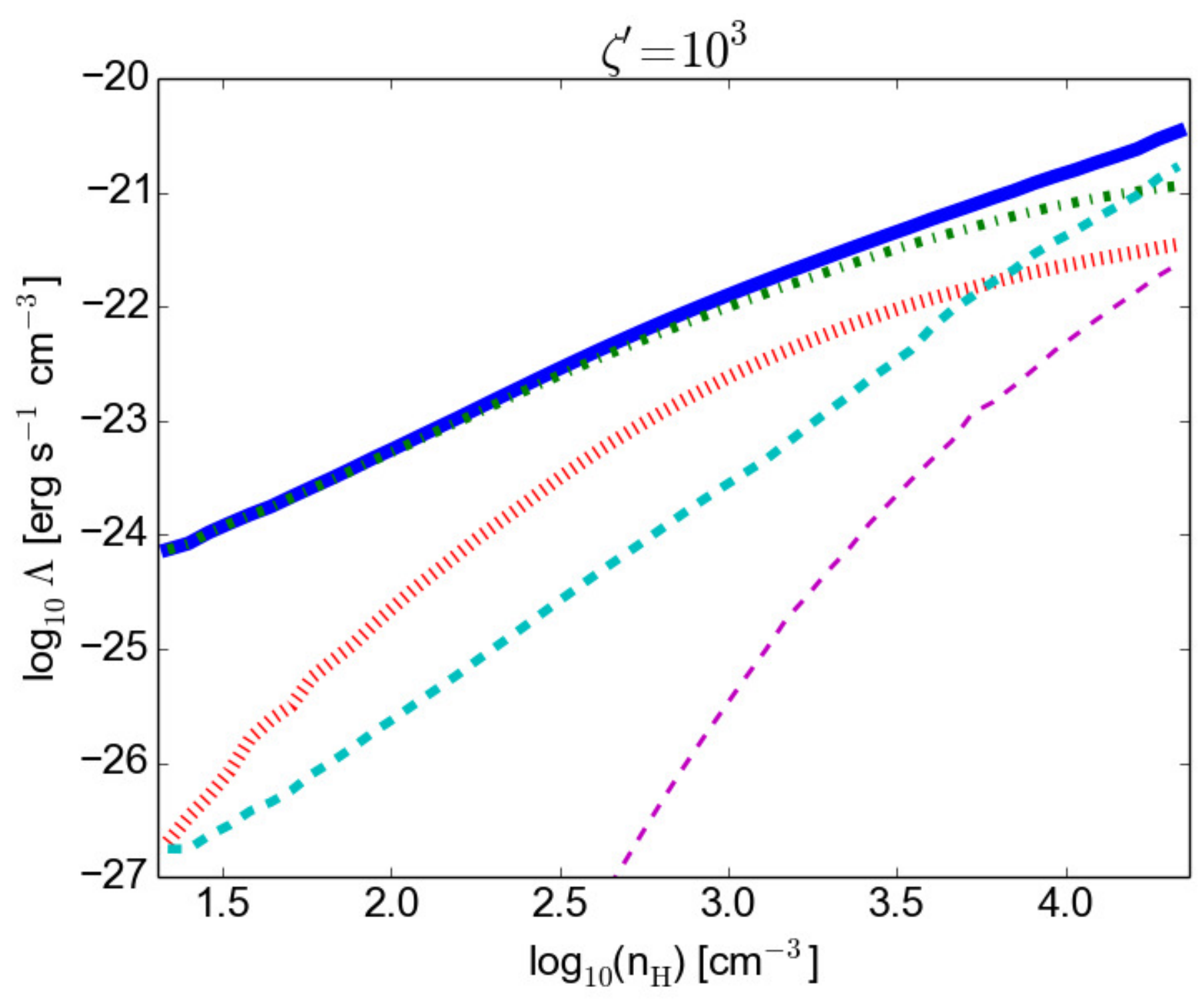}
\caption{ Cooling processes for $\zeta'=1,10^3$. The curves correspond to the mean values for the entire density distribution. At $\zeta'=1$, low density gas is cooling down due to the emission of C$^+$ $158\mu$m fine structure line, intermediate densities due to C whereas gas at higher densities due to CO rotational lines. However, for $\zeta'=10^3$, the reaction with He$^+$ has so severely destroyed CO and partially C that over the entire density distribution the main coolants are C$^+$ up to $\sim3\times10^{3}\,{\rm cm}^{-3}$ and O for higher densities.}
\label{fig:cool}
\end{figure}

\section{Results}
\label{sec:results}

\subsection{Dependency of column density and volumetric mass of species on $\zeta'$}
\label{ssec:3dpdr}

Our description begins with analysing the abundance distribution of species in all four different {\sc 3d-pdr} simulations. \citet{Lada88}, \citet{vanD88} and \citet{vanD92} were the first to divide the gas in `CO-poor' and `CO-rich' populations based on the abundance ratio of [CO]/[H$_2$]. In this work, we adopt the B15 definition for which `CO-deficient' refers to gas that fullfills the conditions
\begin{eqnarray}
{\rm [CO]}/{\rm [H}_2]&<&10^{-5} \label{eqn:co}\\
{\rm [H}{\textsc i}]/2{\rm [H}_2]&<&0.5. \label{eqn:hi}
\end{eqnarray}
In this case, the abundance of CO averaged over the cloud is $\sim\!10\times$ lower than the average value of $\sim10^{-4}$ typically found in molecular clouds while the gas remains H$_2$-rich\footnote{In B15 the gas fulfilling conditions (\ref{eqn:co}) and (\ref{eqn:hi}) was defined as `CO-dark'.}. We define the gas `CO-rich' when the gas is H$_2$-rich and [CO]/[H$_2$]$\geq10^{-5}$. For comparison with observations, it is {\it column density} ratios rather than just local abundance ratios that are most relevant, since column densities ultimately control the strength of the velocity-integrated line emission.

Figure \ref{fig:cd} shows column density plots of H$_2$, H{\sc i}, CO, C, and C$^+$ species as well as cross-section plots of the gas temperature ($T_{\rm gas}$) at the $z=0\,{\rm pc}$ plane, all as a function of $\zeta'$. We map the SPH particle distribution on a $256^3$ grid using the method described in Appendix \ref{app:sph2grid}. The recovery of $T_{\rm gas}\sim10\,{\rm K}$ for the H$_2$ gas inside the cloud with $\chi/\chi_0=1$ and $\zeta'=1$ obtained using our thermal balance calculations is in agreement with the typical temperatures of FUV-shielded H$_2$ observed in our Galaxy \citep[e.g.][]{Poly12,Nish15} and found in other simulations \citep[e.g.][]{Glov12}. On the high end of the average CR energy densities and similar gas densities we recover similar $T_{\rm gas}$ values as in the calculations for CR-dominated regions (CRDRs) performed in the past \citep[][their Figure 1]{Papa11}.

We find that as $\zeta'$ increases, the column density of molecular hydrogen, $N$(H$_2$) remains remarkably unaffected for $\zeta'$ up to $10^{3}$. We note, however, that if we were to evolve the cloud hydrodynamically \citep[e.g.][]{Glov16} the higher gas temperature of the cloud would act to reduce the number of high density clumps, thus affecting the underlying total column density distribution and the chemistry itself. $N$(H{\sc i}) remains low and nearly constant with $\zeta'$ up to $\zeta'$=$10^3$ when H$_2$ starts being significantly destroyed towards H{\sc i}. These trends further reflect the findings of \citet{Bial15} who use the $\zeta_{_{\rm CR}}/n_{\rm H}$ ratio to determine whether the ISM gas is predominantly atomic or molecular. The thin H{\sc i} shell seen in Fig.~\ref{fig:cd} results from photodissociation by the FUV radiation, and the column density $\sim10^{21}\,{\rm cm}^{-2}$ is in agreement with \citet{Ster14} and \citet{Bial16}.

The most interesting interplay is between CO, C and C$^+$. As can be seen from Fig.~\ref{fig:cd}, $N$(CO) starts already decreasing from $\zeta'\simeq10$. For $\zeta'\gtrsim10^2$ it is everywhere approximately one order of magnitude lower than at $\zeta'=1$. Note that for $N$(CO) at $\zeta'=10^2$ and $10^3$ the upper limit of the colour bar is already one order of magnitude less than for $\zeta'=1$ and 10. While $N$(H$_2$) remains high even at $\zeta'\simeq10^3$, the large decrease of $N$(CO) points to a CO-to-H$_2$ conversion factor  well above its Galactic value, and one that may well become uncalibratable (see \S\ref{sec:discussion}). At the same time, as CR particles interact with He, they create He$^+$ ions which then react with CO forming C$^+$. The latter further recombines with free electrons to form neutral carbon. On the other hand $N$(C) increases already from $\zeta'\gtrsim10$, peaking at $\zeta'\sim10^2$. As shown in Fig.~\ref{fig:relations} for the particular comparison between $\zeta'=1$ and $10^2$, it is remarkable to find that
\begin{eqnarray}
N(\textrm{H}_2)_{\zeta'=1}&\simeq&N(\textrm{H}_2)_{\zeta'=100} \label{eqn:rel1} \\
N(\textrm{C}^+)_{\zeta'=1}&\simeq&N(\textrm{C}^+)_{\zeta'=100} \label{eqn:rel2} \\
N(\textrm{C})_{\zeta'=1}&\simeq&N(\textrm{C})_{\zeta'=100} \label{eqn:rel3} \\
N(\textrm{CO})_{\zeta'=1}&\simeq&30N(\textrm{CO})_{\zeta'=100} \label{eqn:rel4}
\end{eqnarray}
suggesting that nearly all CO that has been destroyed by CRs is converted to C$^+$ and C. It is evident from Fig.~\ref{fig:cd} that while at $\zeta'=1$ CO traces the H$_2$ structure very well, it only traces regions of higher column densities at $\zeta'=10$, whereas at $\zeta'=10^2$ it is almost vanished. It is then replaced primarily by C showing a much better resemblance with the molecular structure.

An insidious aspect of a CR-controlled [CO]/[H$_2$] abundance ratio inside CR-irradiated clouds revealed by Fig.~\ref{fig:cd} is that if one were to perform typical CO line observations meant to find the H$_2$ mass and also characterize average gas density and temperature via CO and $^{13}$CO line ratios, their analysis would consistently indicate dense and warm gas, located in those cloud regions where CO manages to survive in a high-CR environment. {\it Yet these routine observations would be totally oblivious to the CO-poor H$_2$ gas mass (and its conditions) that surrounds these CO-rich warm and dense gas `peaks'}. For $\zeta'\geq100$ that would be most of the H$_2$ gas (see Fig.~\ref{fig:cd}), an effect that may have wide-ranging implications for galaxies where most of the SF in the Universe occurs. Apart from dust continuum emission, only C and C$^+$ line imaging could reveal that extra gas mass. Of these two, only C line imaging offers a practical method using ground-based telescopes, since the very high frequency of the C$^+$ line makes it inaccessible for imaging over much of redshift space where star-forming galaxies evolve.

Figure~\ref{fig:masses}  shows the total mass of the above species in the simulated GMC as a function of $\zeta'$. The total mass of H$_2$ in the GMC remains nearly unchanged for up to $\zeta'\sim10^2$. It is expected that for $\zeta'>10^4$ the GMC will be H{\sc i} dominated with only trace amounts of H$_2$ even at the most dense regions. The particular mass of atomic carbon appears to have a local maximum at $\zeta'\sim10^2$, at which point the mass of CO is two orders of magnitude less than the corresponding value for $\zeta'=1$. On the other hand the mass of C$^+$ increases monotonically at all times while for $\zeta'=10^3$ we find that $M$(C$^+$)$\simeq$$M$(C).  

It is interesting to see that the masses of H{\sc i} and C$^+$ increase monotonically, with the mass of C$^+$ increasing somewhat faster than that of H{\sc i}. Both of these species are products of cosmic-rays interacting with H$_2$, CO and C hence it is expected that by increasing $\zeta'$ their abundances will also increase. The observed trend, however, is likely to be a result of additional volumetric (3D) effects.

\begin{figure*}
\center
\includegraphics[width=0.47\textwidth]{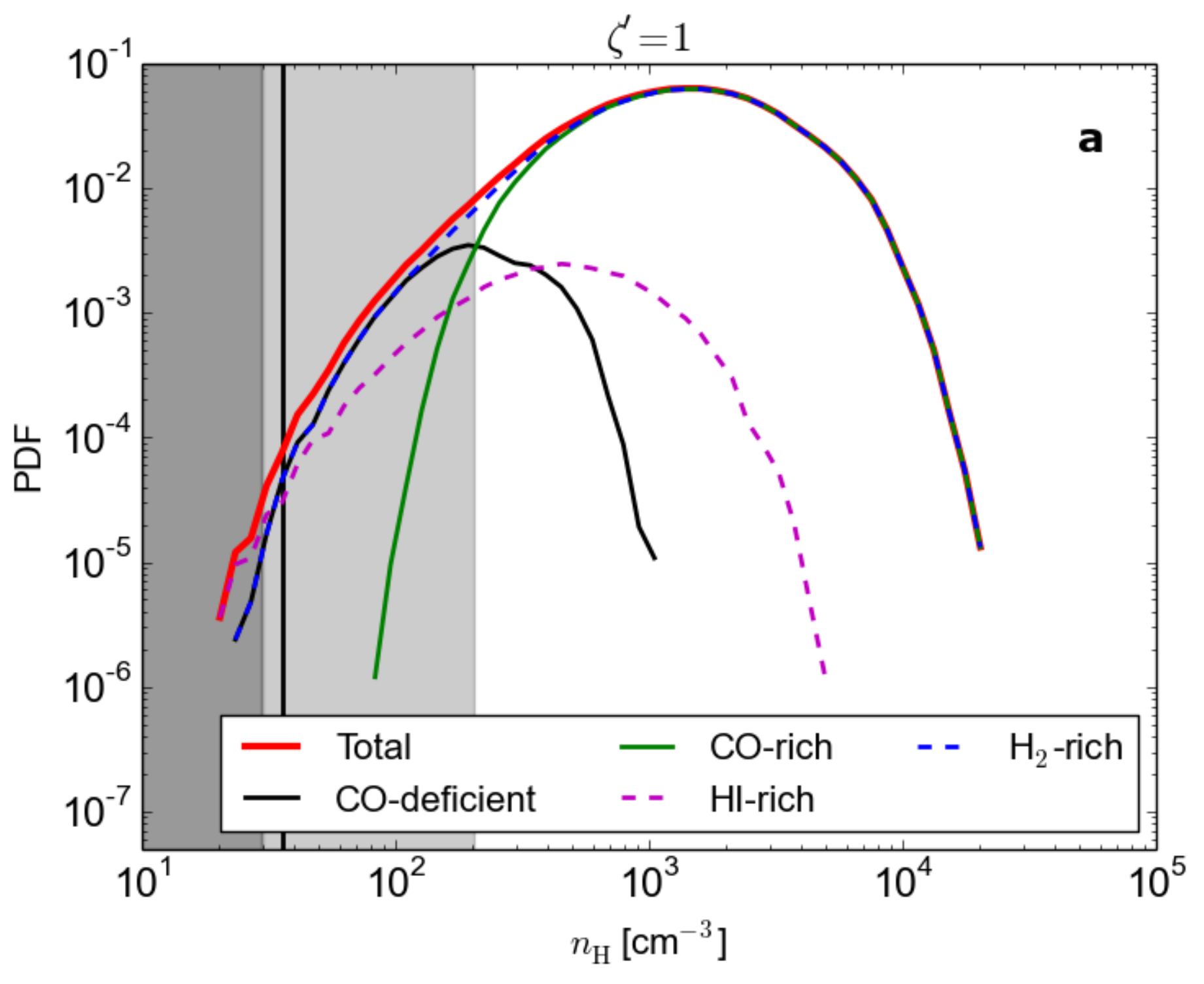}
\includegraphics[width=0.47\textwidth]{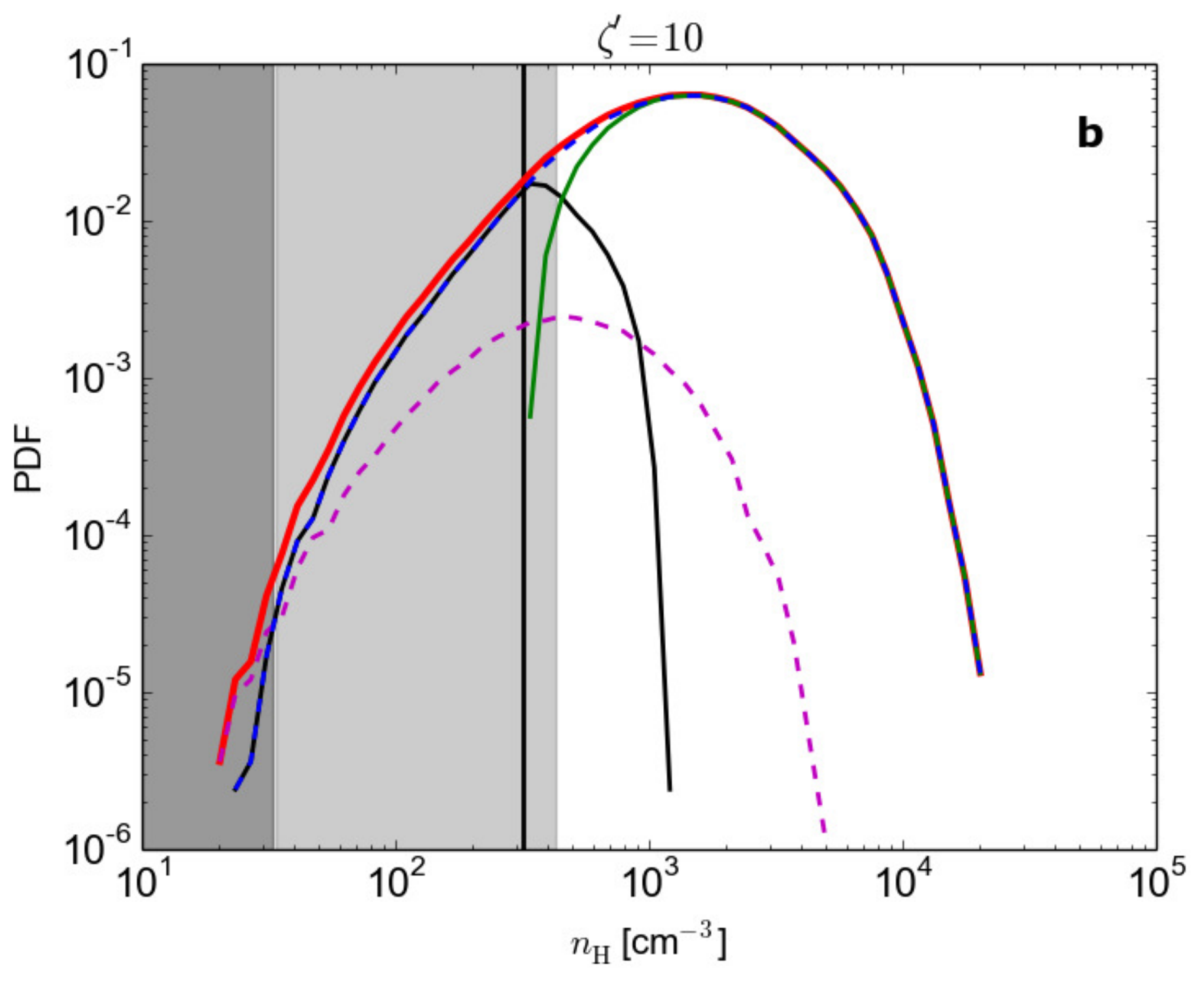}
\includegraphics[width=0.47\textwidth]{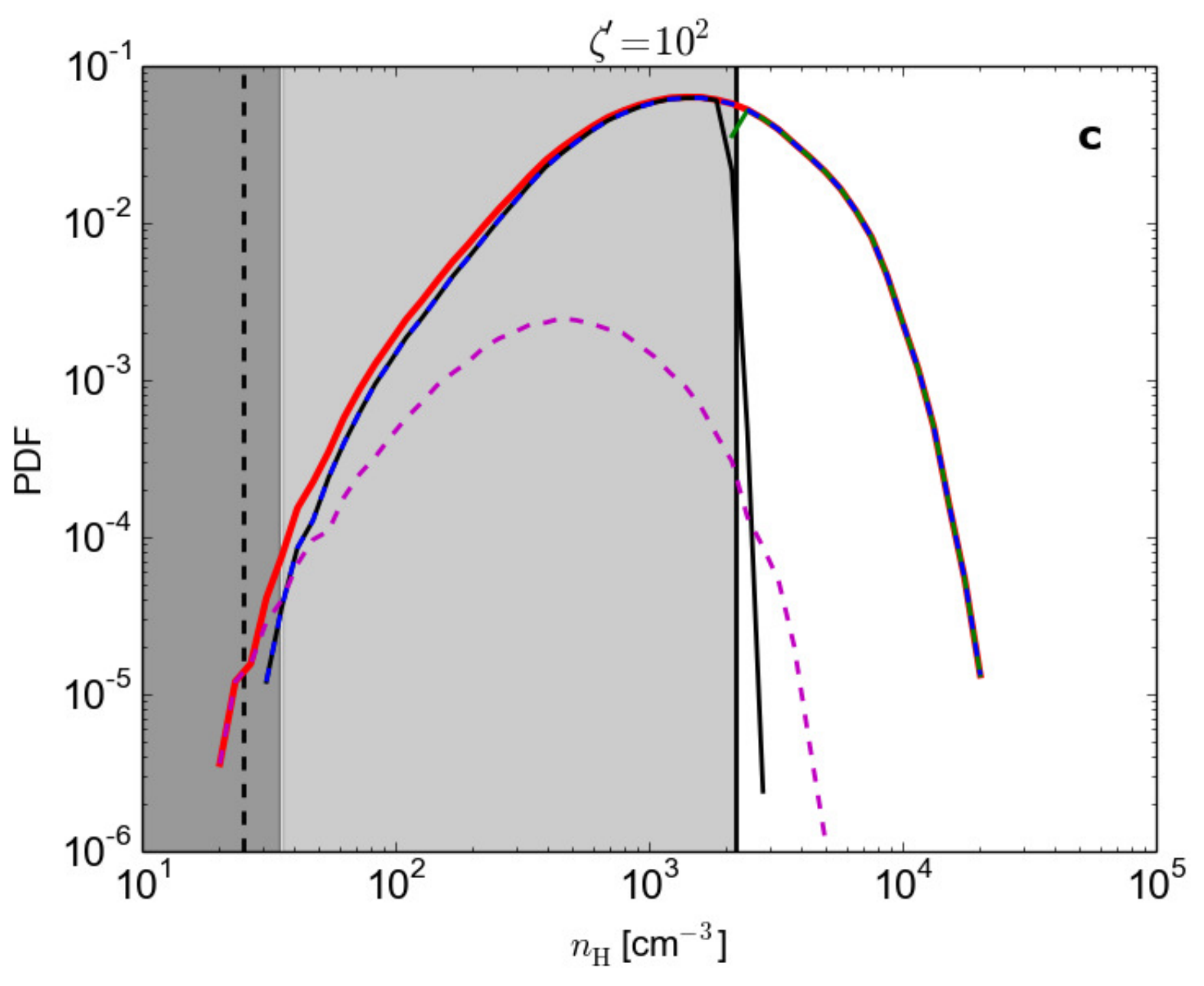}
\includegraphics[width=0.47\textwidth]{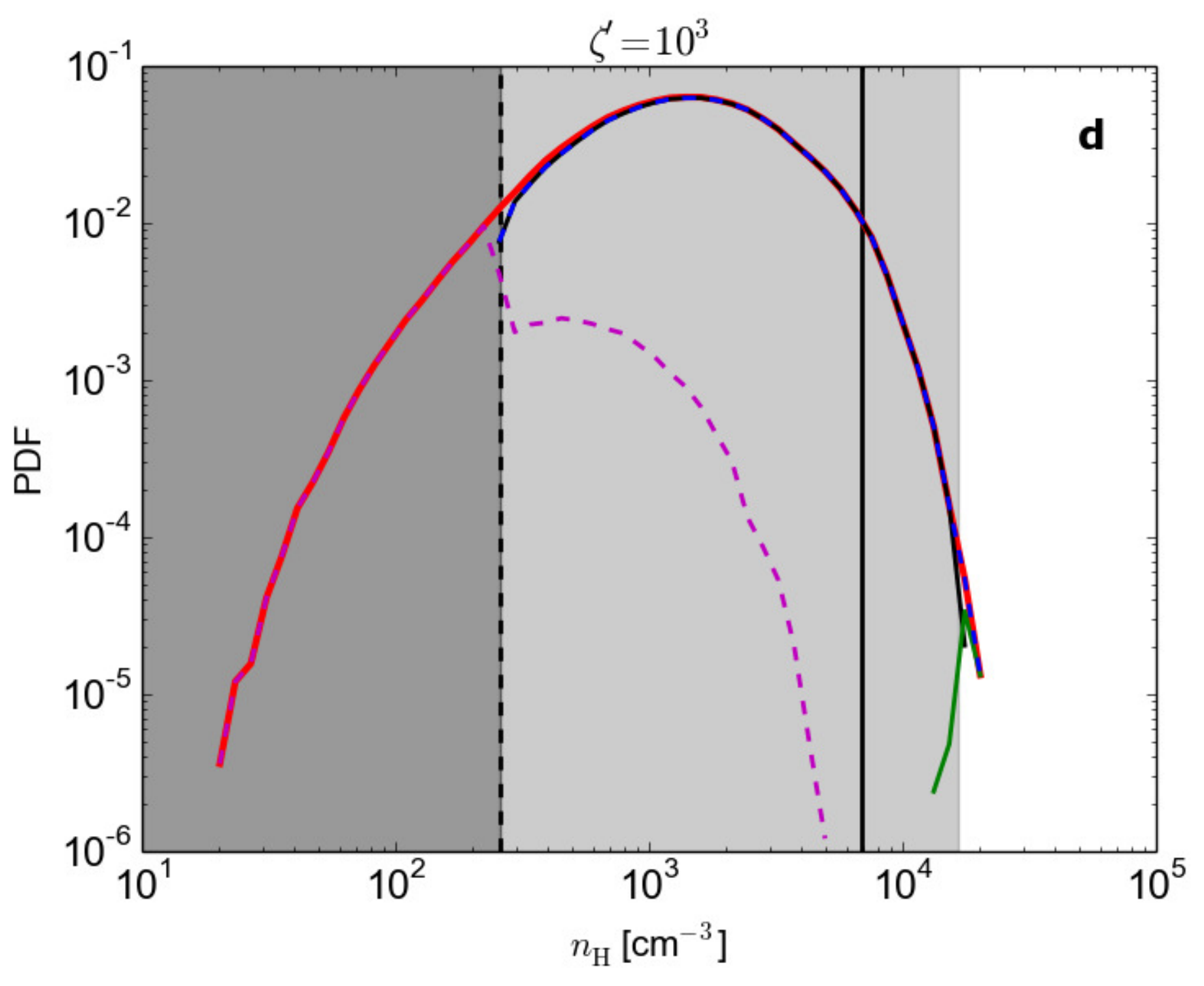}
\caption{ Mass-weighted probability density distribution functions (PDF) of the GMC at different $\zeta'$. In panel (a) $\zeta'=1$, in panel (b) $\zeta'=10$, in panel (c) $\zeta'=10^2$ and in panel (d) $\zeta'=10^3$. The red solid line corresponds to the PDF of the total H-nucleus number density of the SPH particles comprising the cloud. The black line corresponds to the PDF of the `CO-deficient' gas where [CO]/[H$_2$]$<10^{-5}$ and the green line corresponds to PDF of the `CO-rich' gas. The density range of the CO-deficient gas is light-shaded. The PDF blue line corresponds to the H$_2$-rich gas where [H{\sc i}]/2[H$_2$]$<0.5$, whereas magenta is for the H{\sc i}-rich gas. The density range of the latter is dark-shaded. For comparison, vertical lines mark the [CO]/[H$_2$]$=10^{-5}$ (vertical solid) and H{\sc i}/2H$_2=0.5$ (vertical dashed) densities as calculated by B15. Due to the additional photodissociation of CO by the $\chi/\chi_0=1$ FUV isotropic field \citep[normalized according to][]{Drai78}, we find a slightly different density range for the CO-deficient gas at $\zeta'=1$, as such low densities are found at low visual extinction (see Fig.~\ref{fig:aveff_nh}). For $\zeta'>10$ we obtain very good agreement with the prediction of B15.}
\label{fig:pdfs}
\end{figure*}

\subsection{Heating and cooling processes}
\label{ssec:heatcool}

The {\sc 3d-pdr} code performs thermal balance iterations and converges when the total heating rate matches with the total cooling rate calculated for each position within the cloud. The heating processes considered include the \citet{Bake94} grain and PAH photoeletric heating with the modifications suggested by \citet{Wolf03} to account for the revised PAH abundance estimate from the {\it Spitzer} data; carbon photoionization heating \citep{Blac87}; H$_2$ formation and photodissociation heating \citep{Tiel85}; collisional de-excitation of vibrationally excited H$_2$ following FUV pumping \citep{Holl79}; heat deposition per cosmic-ray ionization \citep{Tiel85}; supersonic turbulent decay heating \citep{Blac87}; exothermic chemical reaction heating \citep{Clav78}; and gas-grain collisional coupling \citep{Burk83}. The particular H$_2$ formation rate is calculated using the treatment of \citet{Caza02,Caza04}. The turbulent heating that is included in {\sc 3d-pdr} is $\propto v_{\rm turb}^3/L$, where $v_{\rm turb}=1.5\,{\rm km}\,{\rm s}^{-1}$ and $L=5\,{\rm pc}$. These values are constant throughout all calculations, giving $v_{\rm turb}^3/L\sim2\times10^{-4}\,{\rm cm}^2{\rm s}^{-3}$. Our chosen $v_{\rm turb}$ is on par of what is expected from the Larson relation and its subsequent observational study by \citet{Solo87}. This turbulent heating term assumes that turbulence is driven at the largest scale of the cloud \citep{Heye04,Pado09}. 

The gas primarily cools due to the collisional excitation, subsequent C$^+$, C, O fine-structure line emission and emission due to rotational transitions of CO. The cooling rate of each processes are estimated using a 3D escape probability routine. The details are described in \citet{Bisb12} and the data-files are adopted from the Leiden Atomic and Molecular Database \citep[LAMDA][]{Scho05}\footnote{http://home.strw.leidenuniv.nl/$\sim$moldata/}. We use a macroturbulent expression to account for the optical depth \citep[][and Appendix A of B15]{Papa99}.

For densities $n_{_{\rm H}}\lesssim10^2\,{\rm cm}^{-3}$ located mainly at the outer regions of the GMC, heating comes predominantly from photoelectrons which are produced due to the isotropic FUV radiation field (see Fig.~\ref{fig:heat}). For $10^2\lesssim n_{_{\rm H}}\lesssim10^3\,{\rm cm}^{-3}$ and for all $\zeta'$, we find that heating results predominantly from contributions due to photoelectrons, dissipation of turbulence, exothermic reactions due to recombinations of HCO$^+$, H$_3^+$, H$_3$O$^+$ and ion-neutral reactions of He$^+$ + H$_2$ (chemical heating), energy deposition due to cosmic-ray reactions and heating due to H$_2$ formation. For higher densities and for $\zeta'=1$, heating results from the turbulence with smaller contributions from cosmic rays and chemical heating. As $\zeta'$ increases, however, we find that cosmic rays dominate over all other heating mechanisms. The chemical heating also contributes significantly. The latter results from the abundance increase of all participating ions due to reactions ignited by the high cosmic-ray energy density. This is reflected in the lower panel of Fig.~\ref{fig:heat} where we show the heating functions at $\zeta'=10^3$.

Likewise, cooling depends on $n_{_{\rm H}}$ and $\zeta'$ (see Fig.~\ref{fig:cool}). In particular, for all $\zeta'$ we find that at low densities, cooling results predominantly from C$^+$ which -- along with photoelectric heating -- controls the gas temperature at the outer shell of the GMC. The increase of the cosmic-ray ionization rate results in the increase of C$^+$ abundance and hence its cooling efficiency, which in turn decreases the gas temperature. This is actually the reason why the gas temperature is lower at low A$_{\rm V,eff}$ (see \S\ref{ssec:pdfs}) with increasing cosmic-rays (see \S\ref{ssec:tgas}). This result has been further reproduced by 1D calculations confirming the importance of the [C$^+$] increase. For $\zeta'\sim10^3$, we find that C$^+$ cooling dominates for densities up to $n_{_{\rm H}}\sim10^3\,{\rm cm}^{-3}$. On the other hand, cooling due to C is important for $\zeta'\lesssim10^2$ particularly for densities between $10^2\lesssim n_{_{\rm H}}\lesssim10^{3.5}$. Finally, for $n_{_{\rm H}}\gtrsim10^{3.5}\,{\rm cm}^{-3}$ CO rotational lines contribute predominantly in the gas cooling for $\zeta'\lesssim10^2$ with O to become substantially important for high densities ($n_{_{\rm H}}>10^{3.5}$) and high cosmic-ray ionization rates ($\zeta'\sim10^3$), although it is not a main coolant in all other cases.

Dust temperatures are calculated for each SPH particle using the treatment of \citet{Holl91} for the heating due to the incident FUV photons. This approach is further modified to include the attenuation of the IR radiation as described by \citet{Rowa80}. Since the UV radiation on the surface of the cloud is approximately $1/4$ Draine (see \S\ref{ssec:ics}), the maximum dust temperature we find is $T_{\rm dust}\sim12\,{\rm K}$, located at large radii. We impose a floor dust temperature of $10\,{\rm K}$ which is consisted with the average lowest temperatures observed \citep[e.g.][]{Plan16}. We can therefore assume that the dust temperature in the entire cloud is approximately uniform and equal to $T_{\rm dust}=10\,{\rm K}$. 

In regions with densities exceeding $10^4\,{\rm cm}^{-3}$, CO freeze-out on dust grains may become an important process and the CO abundance in gas phase can be sufficiently reduced affecting its emissivity. Cosmic-ray induced (photo-)desorption can then bring a small fraction of this gas back to gas phase. Our results would not be altered if we were to include this process. This is because only $\sim 0.4\%$ of the total mass of the simulated cloud has densities exceeding $10^4\,{\rm cm}^{-3}$ and the corresponding CO abundance never exceeds $\sim 16\%$ of the total CO abundance throughout the cloud (for $\zeta'=10^2$ whereas for all other cases it is well below $\sim\!10\%$). Moreover in GMCs typically only small H$_2$ gas mass fractions reside at regions with $n_{\rm H}>10^4\,{\rm cm}^{-3}$ making CO-freeze of little importance for the bulk of their mass.

\subsection{Probability density functions}
\label{ssec:pdfs}

Figure \ref{fig:pdfs} shows mass-weighted probability density distribution functions (PDFs) for each simulation. In these plots it can be seen how the effect of CO-destruction operates volumetrically, particularly when applying conditions (\ref{eqn:co}) and (\ref{eqn:hi}). In all plots, the non-shaded part corresponds to CO-rich densities, the light-shaded part to all H$_2$-rich but CO-deficient densities and the dark-shaded part to all H{\sc i}-rich densities. It is interesting to compare the CO-rich, CO-deficient and H{\sc i} regimes with those predicted by B15 from one-dimensional calculations. For this purpose, we plot also in each case the limits for the CO-deficient (vertical solid) and H{\sc i}-rich (vertical dashed) regions as indicated in the B15 parameter plot (their Fig.~1). For $\zeta'=1$, B15 find that for densities $n_{_{\rm H}}\lesssim25\,{\rm cm}^{-3}$ the gas will be CO-deficient. However, in our 3D simulations we find that for densities up to this value the gas will also be in H{\sc i} form.  The CO-deficient/H$_2$-rich density range now lies in $25\lesssim n_{_{\rm H}}\lesssim200\,{\rm cm}^{-3}$ (Fig.~\ref{fig:pdfs}a). This difference occurs because of the additional effect of photodissociation of CO due to the isotropic FUV radiation being more effective at lower densities which are located at the outer parts of the cloud (larger radii). This radiation creates also some additional amount of H{\sc i} at the outer shell of the cloud, on top of the CR interaction, due to photodissociation of H$_2$.

\begin{figure}
\center
\includegraphics[width=0.45\textwidth]{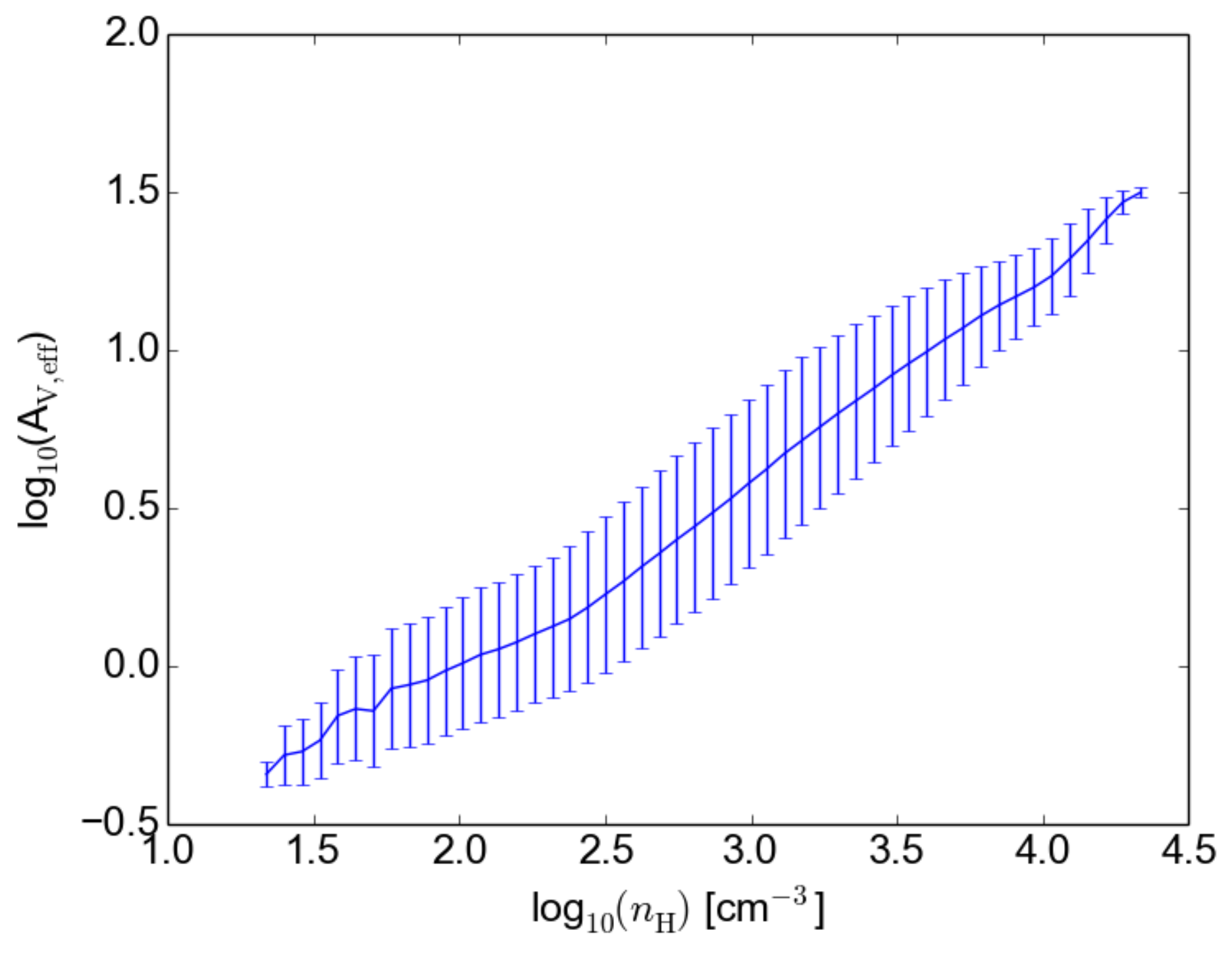}
\caption{ Correlation of  $n_{_{\rm H}}$ with A$_{\rm V,eff}$ for the present GMC. The error-bars correspond to $1\sigma$ standard deviation. Low densities have a low effective visual extinction as they are primarily located at larger radii (cloud edge) whereas high densities are well shielded from the isotropic FUV radiation field.}
\label{fig:aveff_nh}
\end{figure}

The fact that lower densities are located mostly at the outer parts of the cloud is verified in Fig.~\ref{fig:aveff_nh} where we correlate the effective visual extinction \citep[e.g.][]{Glov10,Offn13}, $A_{\rm V,eff}$, defined as
\begin{eqnarray}
\label{eqn:aveff}
{\rm A_{V,eff}}=-0.4\ln\left(\frac{1}{{\cal N}_{\ell}}\sum_{i=1}^{{\cal N}_{\ell}}e^{-2.5{\rm A_V}[i]}\right),
\end{eqnarray}
with the $n_{_{\rm H}}$ number density. This ${\rm A_{V,eff}}$ is different from the observed visual extinction. When looking towards the centre of a spherically symmetric cloud, this expression would give half of the observed ${\rm A_{V}}$, which is calculated from one edge of the cloud to the other. In the above equation, ${\cal N}_{\ell}$ corresponds to the number of HEALPix \citep{Gors05} rays we used and which is equal\footnote{Following the analysis by \citet{Offn13} we do not expect our result to sensitively depend on the chosen angular resolution. See also \citet{Clar12}} to 12. Indeed from Fig.~\ref{fig:aveff_nh} we find that densities of $n_{\rm H}\lesssim200\,{\rm cm}^{-3}$ have mean visual extinction of A$_{\rm V,eff}\lesssim0.8\,{\rm mag}$ and are located mainly at the outer shell of the cloud \citep[see also][]{Wu15}. They are therefore affected by the FUV radiation.

For $\zeta'=10$ and $10^2$ (Fig.~\ref{fig:pdfs}b, c) we find very good agreement with the B15 parameter plot in estimating the density range of the CO-deficient gas. As discussed above, this is the range of cosmic-ray ionization rates for which we obtain high abundances of C while the gas remains almost entirely H$_2$-rich. As can be seen in both cases, the [CO]/[H$_2$] ratio is $\gtrsim10^{-5}$  only for moderate/high densities \citep[i.e. $n_{_{\rm H}}\gtrsim500\,{\rm cm}^{-3}$ and $n_{_{\rm H}}\gtrsim2\times10^3\,{\rm cm}^{-3}$ respectively, see][for an analytical description concerning the dependency of the {[}CO{]}/{[}H$_2${]} ratio as a function of $\zeta_{_{\rm CR}}/n_{\rm H}$]{Bial15}

As seen in Fig.~\ref{fig:pdfs}d, for $\zeta'=10^3$ we find that the density range dominated by H{\sc i} is in agreement with the B15 parameter plot at a remarkable precision. However, the density range of the CO-deficient gas is now wider. Although for this $\zeta'$, B15 predict that the CO-deficient gas will be observed in the $300\lesssim n_{\rm H}\lesssim7\times10^3\,{\rm cm}^{-3}$ density range, the corresponding upper limit that we find here is $\sim1.5\times10^4\,{\rm cm}^{-3}$. As discussed in B15, the `turnover' point is sensitive to the gas temperature obtained from the thermal balance, whereas the latter is also sensitive to the cooling functions, which depend on the density distribution. We therefore assign this discrepancy to the additional 3D effects that cannot be modelled with corresponding 1D calculations.  

\subsection{Abundances distribution and gas temperatures}
\label{ssec:tgas}

\begin{figure}
\center
\includegraphics[width=0.45\textwidth]{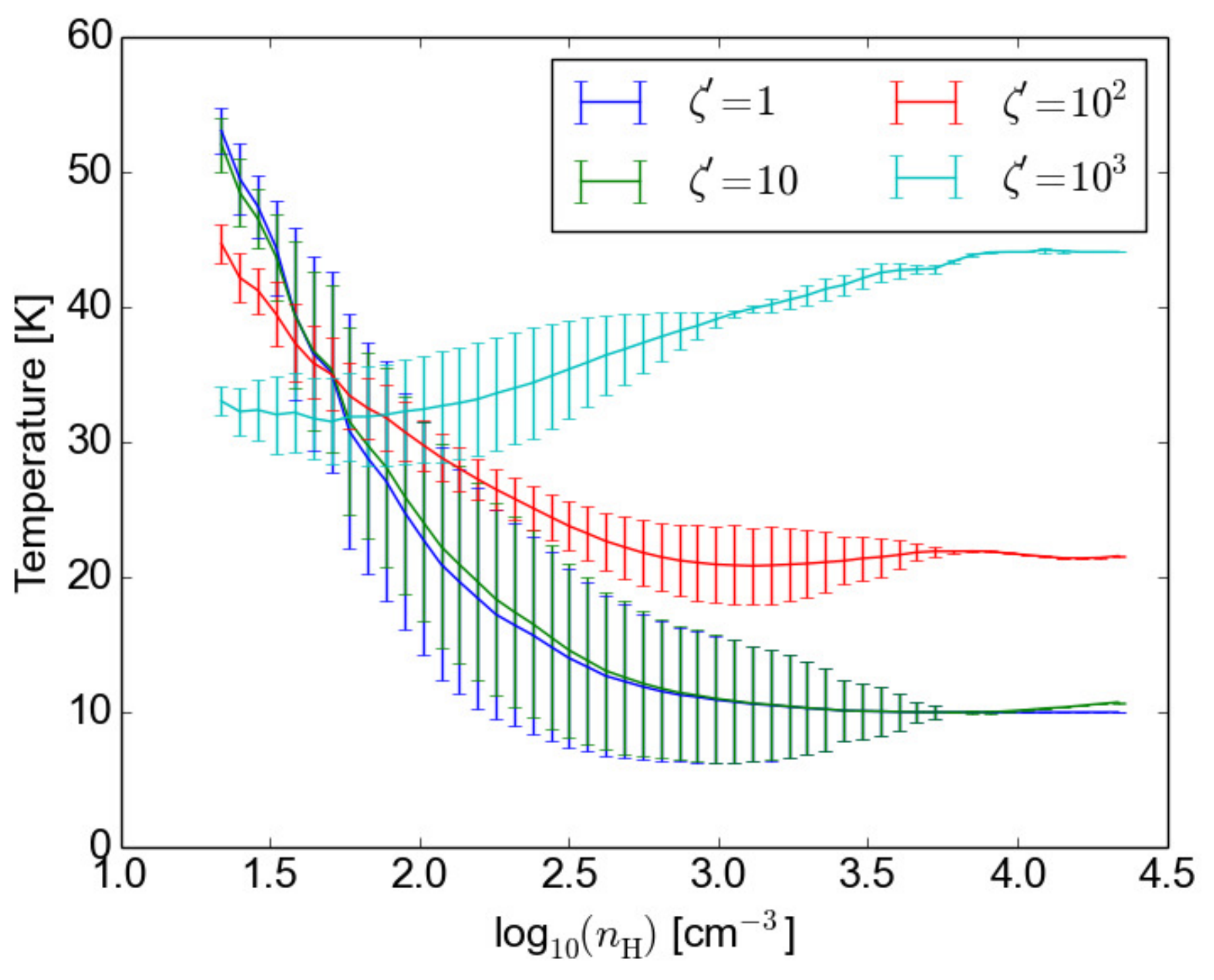}
\caption{ Correlation of $n_{_{\rm H}}$ with the gas temperature, $T_{\rm gas}$. The error-bars correspond to $1\sigma$ standard deviation and they are decreasing with $n_{_{\rm H}}$ since in this regime, chemistry is primarily controlled by cosmic-rays which depends weakly on $n_{_{\rm H}}$. The temperatures obtained for $n_{_{\rm H}}\gtrsim500\,{\rm cm}^{-3}$ are similar to those predicted by B15 (their Fig.~9). For low $n_{_{\rm H}}$ (which are in principle located at low A$_{\rm V,eff}$; see Fig.~\ref{fig:aveff_nh}) we find that $T_{\rm gas}$ decreases by increasing $\zeta'$ as a result of the increase of [C$^+$] abundance and the associated C$^+$ $158\,\mu$m cooling line emission.}
\label{fig:n_tgas}
\end{figure}

In Fig.~\ref{fig:n_tgas} we show the gas temperature, $T_{\rm gas}$, versus the $n_{_{\rm H}}$ number density for all four different $\zeta'$ simulations. For all $\zeta'$ and for $\log_{10}(n_{_{\rm H}})\gtrsim3.5$ we find very good agreement with the predicted $T_{\rm gas}$ of B15 (their Fig.~9). This is because for this range of densities, A$_{\rm V,eff}\gtrsim2\,{\rm mag}$ thus the isotropic FUV is sufficiently attenuated. Note that the standard-deviation bars of $T_{\rm gas}$ at $n_{\rm H}\gtrsim10^3\,{\rm cm}^{-3}$ decrease while for $n_{\rm H}\sim10^4\,{\rm cm}^{-3}$ they are negligible. While the FUV has been severely extinguished, this regime is predominantly controlled by the cosmic-ray interaction which in turn depends very weakly on $n_{\rm H}$ as illustrated in Fig.~9 of B15.
We also find that the mean gas temperatures, $\langle T_{\rm gas}\rangle_{\zeta'}$, in each different $\zeta'$ are $\langle T_{\rm gas}\rangle_1\simeq11\,{\rm K}$, $\langle T_{\rm gas}\rangle_{10}\simeq11\,{\rm K}$, $\langle T_{\rm gas}\rangle_{100}\simeq22\,{\rm K}$ and $\langle T_{\rm gas}\rangle_{1000}\simeq40\,{\rm K}$. The low temperatures obtained for Galactic average CR energy densities are similar to those observed for FUV-shielded dark cores \citep[see][for a review]{Berg07}. Moreover, the fact that $T_{\rm gas}$ remains low and nearly constant for modestly boosted CR energy densities (e.g. $\sim(1-10)\times$Galactic) recovers the result obtained by \citet{Papa11} for uniform clouds, further demonstrating the robustness of the initial conditions of star-formation set deep inside such FUV-shielded dense gas regions. This robustness is an important starting point for all gravoturbulent theories of star-formation inside GMCs \citep[][and references therein]{Papa11}.

Note also that for low densities i.e. $n_{_{\rm H}}<10^2\,{\rm cm}^{-3}$, found mostly in outer cloud layers and in principle exposed to the isotropic FUV radiation, $T_{\rm gas}$ decreases as $\zeta'$ increases. This is because the FUV radiation along with the high CR ionization rate creates large amounts of C$^+$, whose emission line is an effective coolant (as discussed in \S\ref{ssec:heatcool}), driving the decrease of $T_{\rm gas}$.

\begin{figure*}
\center
\includegraphics[width=0.32\textwidth]{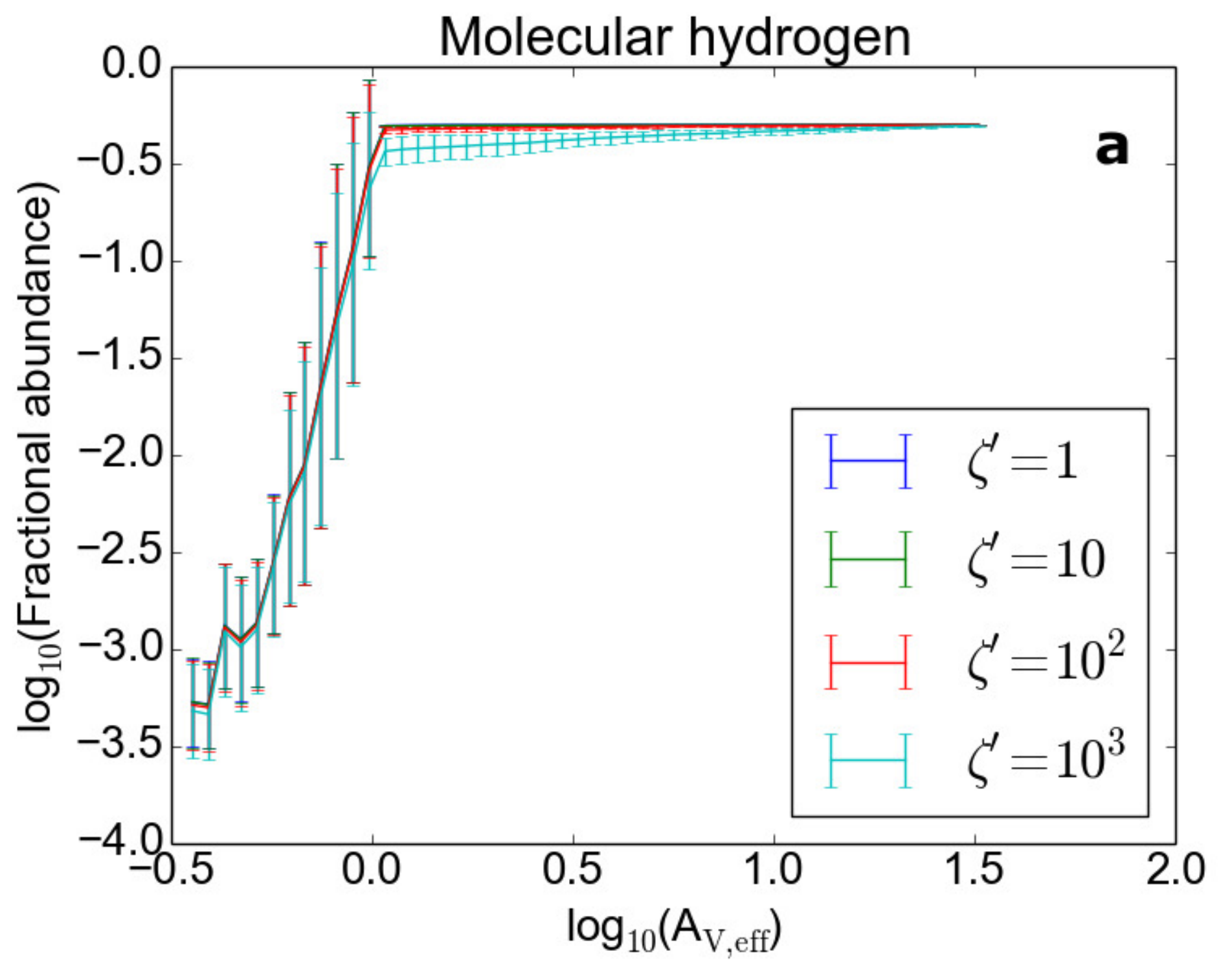}
\includegraphics[width=0.32\textwidth]{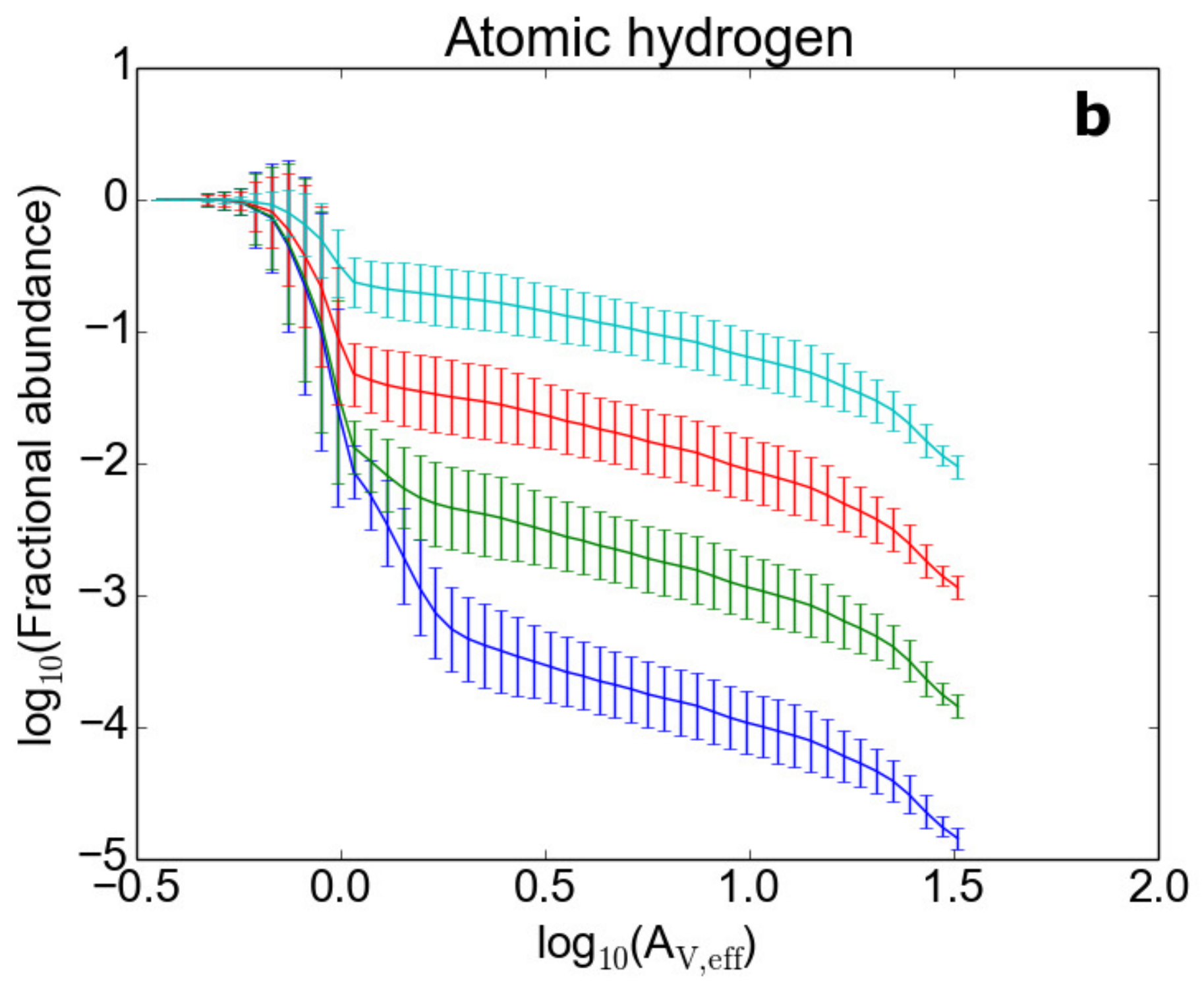}
\includegraphics[width=0.32\textwidth]{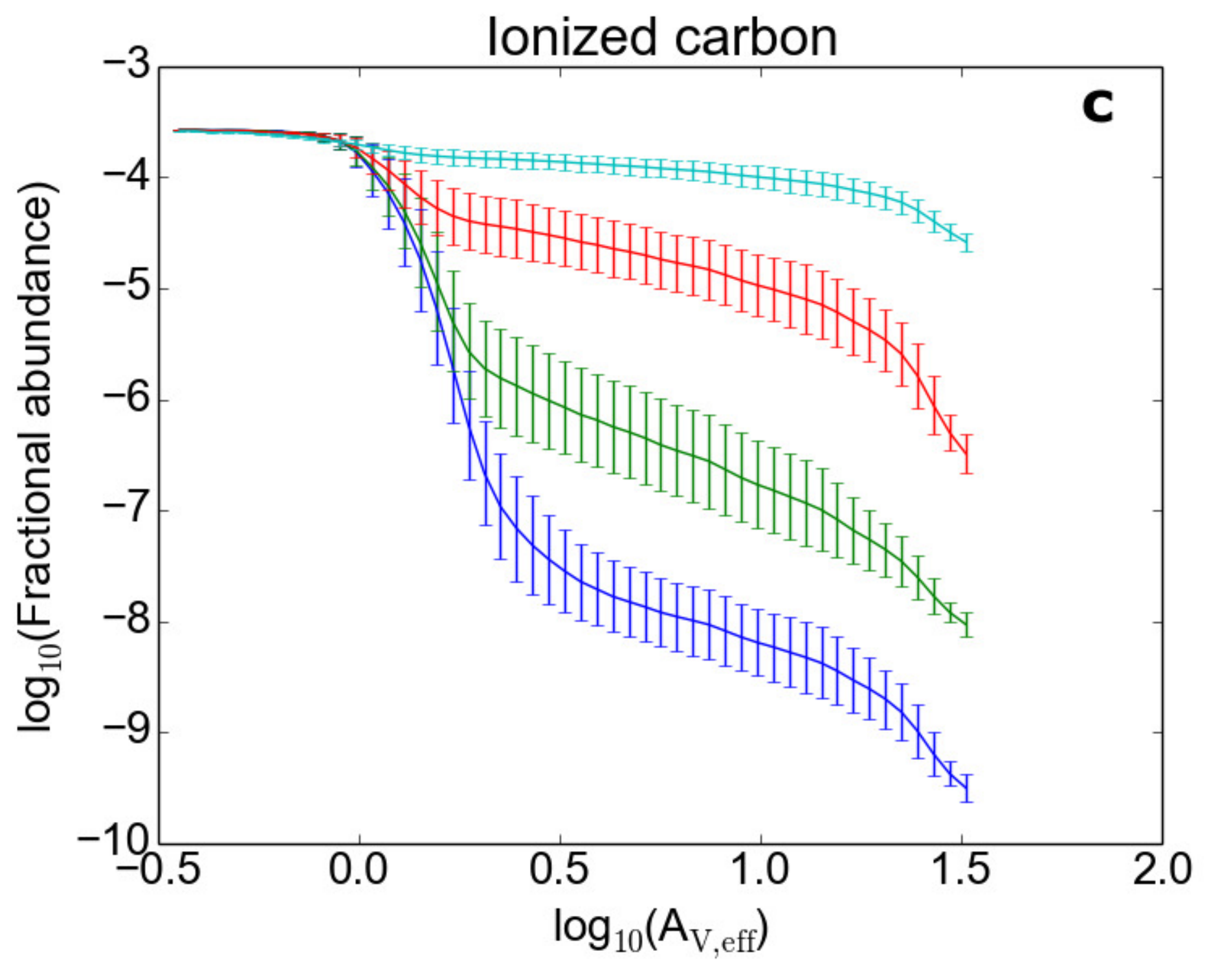}
\includegraphics[width=0.32\textwidth]{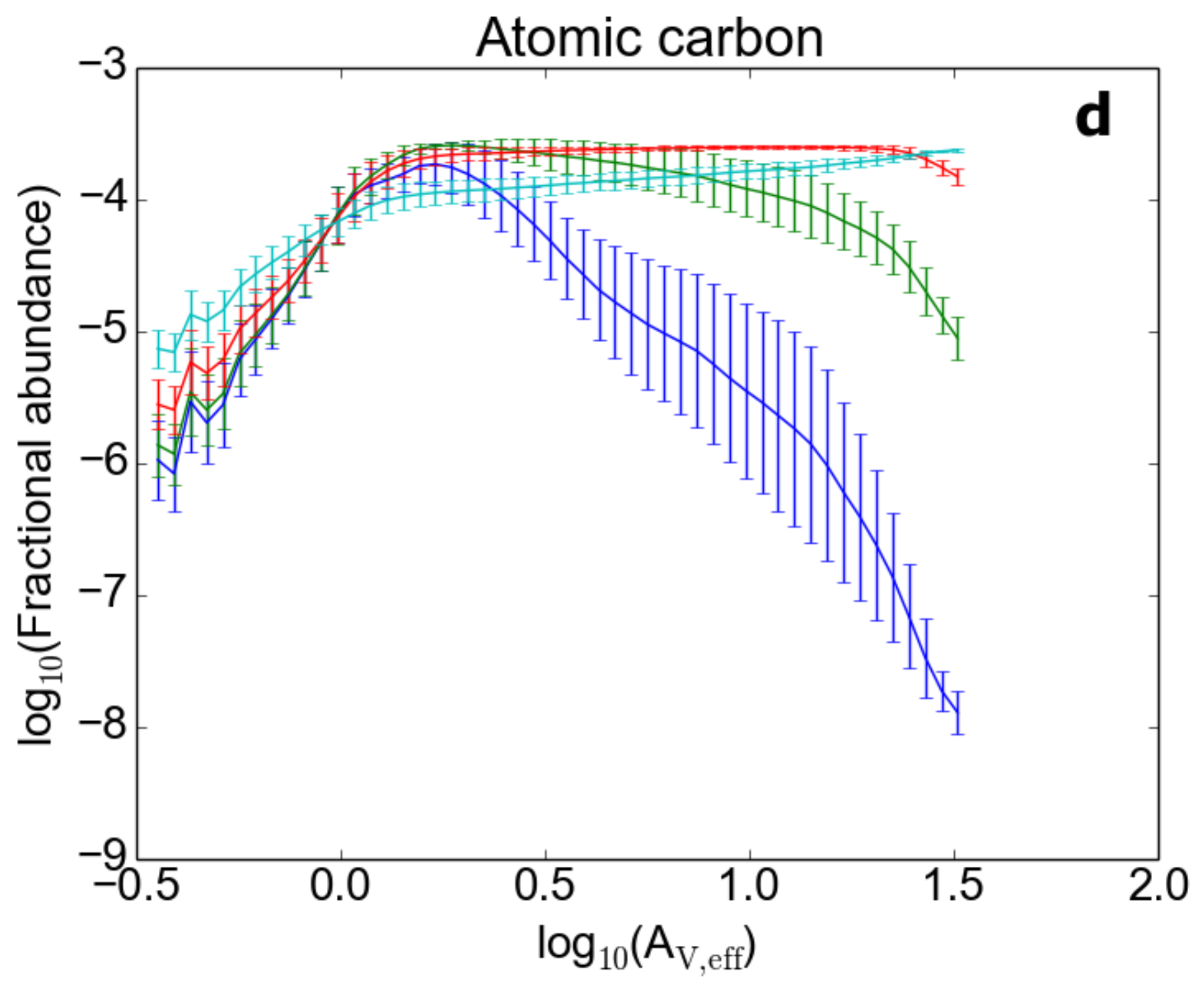}
\includegraphics[width=0.32\textwidth]{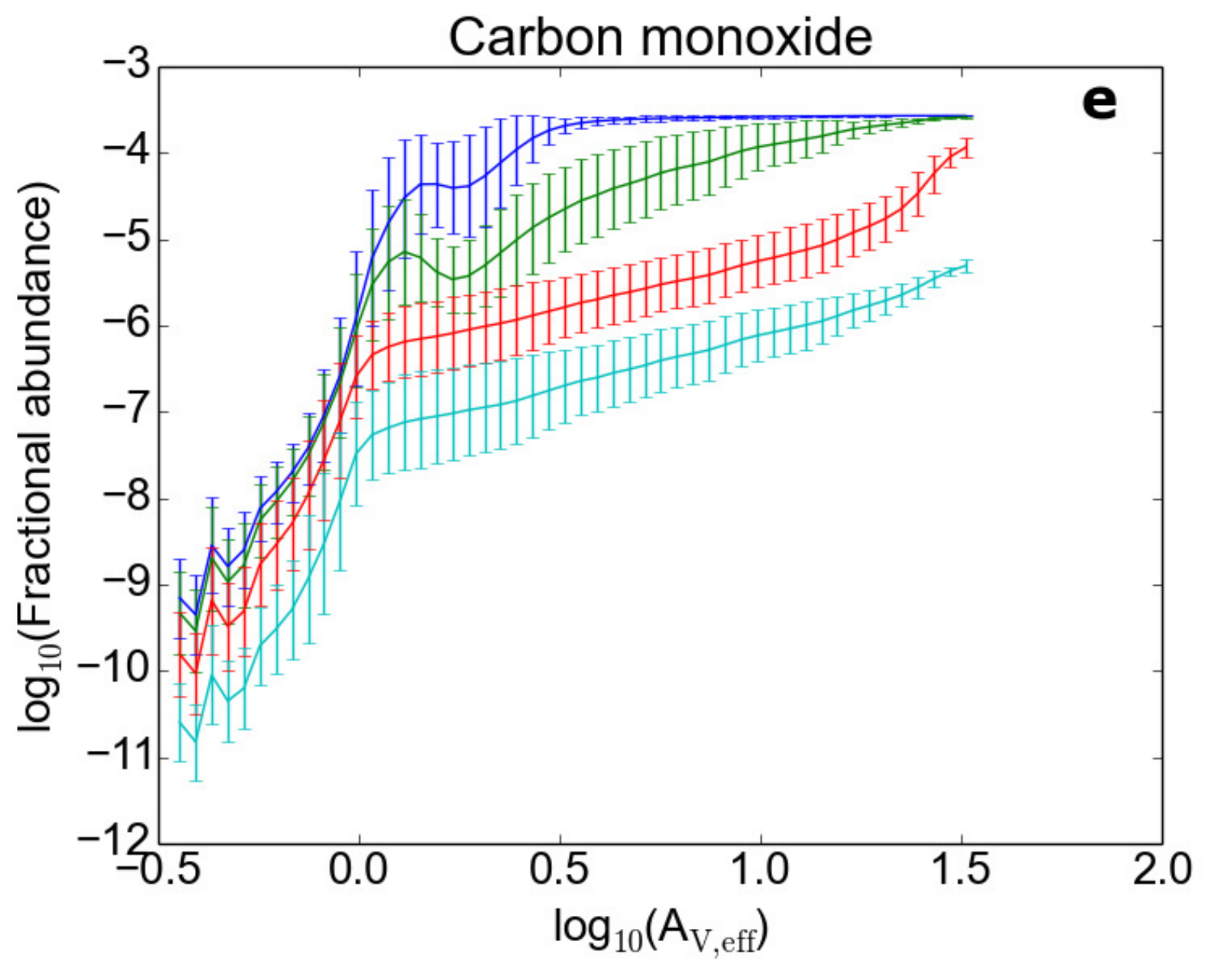}
\includegraphics[width=0.32\textwidth]{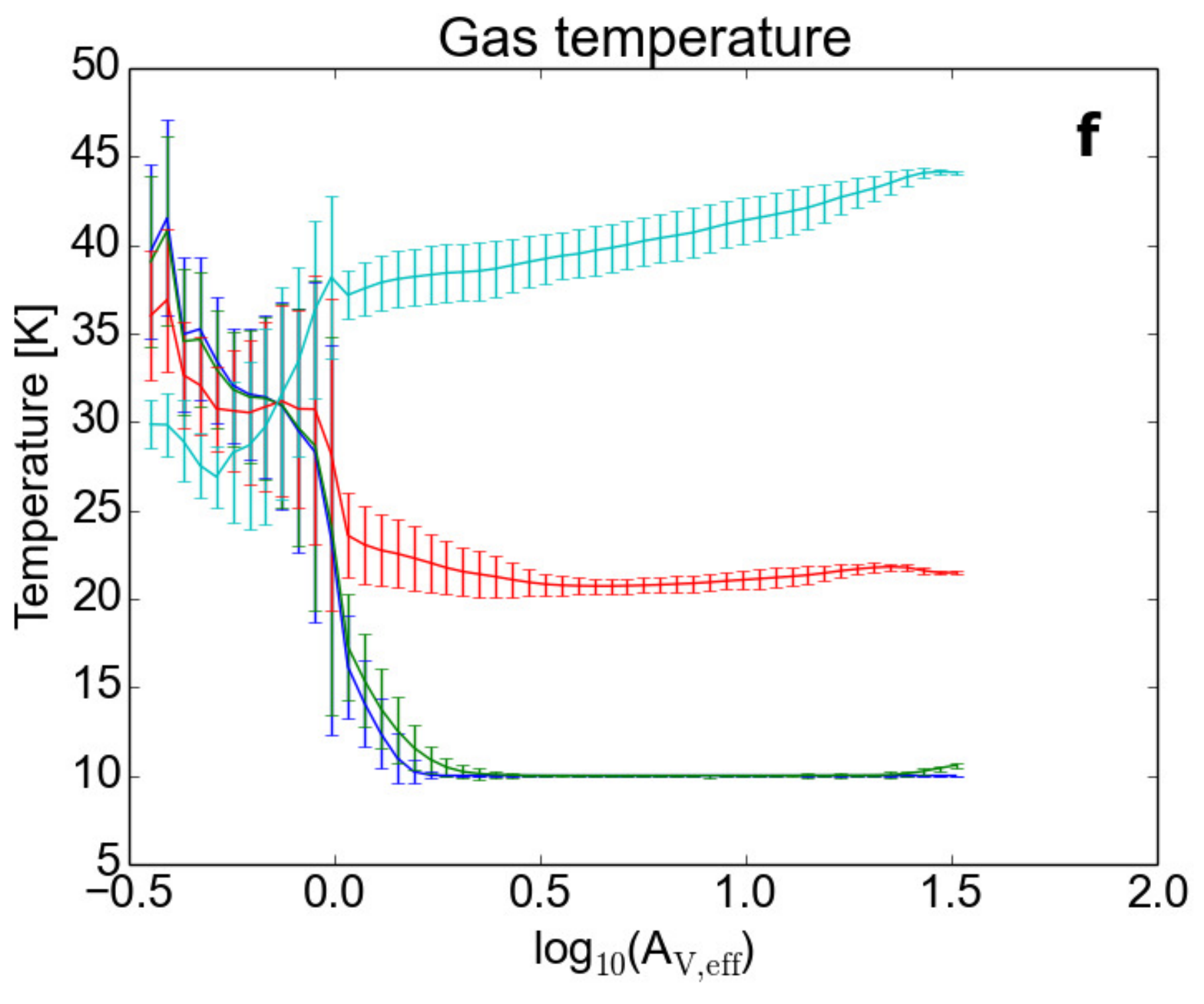}
\caption{ Correlation of A$_{\rm V,eff}$ (see Eqn.~\ref{eqn:aveff}) with the fractional abundances of H$_2$ (panel a), H{\sc i} (panel b), C$^+$ (panel c), C (panel d), CO (panel e) and $T_{\rm gas}$ (panel f) for all four different $\zeta'$. The error-bars correspond to $1\sigma$ standard deviation and they are decreasing at higher A$_{\rm V,eff}$ since in this regime, chemistry is primarily controlled by cosmic-rays which depends very weakly on $n_{_{\rm H}}$. The H$_2$ fractional abundance remains remarkably similar in all different $\zeta'$ whereas CO is destroyed by CR particles forming C$^+$ and C. Observe also that in panel f and at $\log_{10}({\rm A_{V,eff}})\lesssim-0.3$, $T_{\rm gas}$ decreases with increasing $\zeta'$ by $\sim10\,{\rm K}$ when compared to the two $\zeta'$ extrema (see also Fig.~\ref{fig:n_tgas}). It can further be seen that for $\zeta'\gtrsim10^2$ the gas temperature at high visual extinction is approximately uniform and entirely controlled from heating mechanisms caused by cosmic rays.}
\label{fig:aveff}
\end{figure*}

To further understand  how the abundance distribution of species changes with $\zeta'$ it is convenient to correlate them with A$_{\rm V,eff}$. This is shown in Fig.~\ref{fig:aveff} where panels (a)-(e) show the abundances of H$_2$, H{\sc i}, C$^+$, C and CO, and panel (f) that for $T_{\rm gas}$ versus A$_{\rm V,eff}$. As demonstrated earlier, the abundance of H$_2$ remains remarkably similar under all $\zeta'$ values. The differences in H$_2$ abundance as a function of $\zeta'$ are reflected in the abundance of H{\sc i} in each case; here we can see that for all $\zeta'$, [H{\sc i}]$\lesssim10^{-1}$ in the interior of the cloud i.e. where A$_{\rm V,eff}>7\,{\rm mag}$. On the contrary, C$^+$, C and CO depend more sensitively on an increasing $\zeta'$ with CO abundance destroyed even at high density clumps close to the centre of the GMC when $\zeta'$ are high. As expected, C and C$^+$ follow the reverse trend in which they increase in abundance with increasing $\zeta'$. Observe again that for $\zeta'=10^3$ the abundance of C is less than in $\zeta'=10^2$ (as it is also destroyed) indicating that there is a range of cosmic-ray energy densities for which the overall abundance of C peaks and where the C-to-H$_2$ method will be particularly robust. Note that in both Fig.~\ref{fig:n_tgas} and Fig.~\ref{fig:aveff}f, the error bars (corresponding to $1\sigma$ standard deviation) are much smaller for high $n_{\rm H}$ and A$_{\rm V,eff}$ respectively meaning that $T_{\rm gas}$ in this regime is approximately uniform and entirely controlled by cosmic-ray heating.

\section{Thermal Balance and the crucial role of OH}
\label{sec:oh}

The CO molecule can form through various channels \citep[e.g.][]{Herb73,vanD88,Ster95,Tiel13}. An important formation route, especially at moderate-to-high cosmic-ray or X-ray ionization rates, as well as in low metallicity gas \citep{Bial15}, depends on the OH intermediary. In cold gas ($T_{\rm gas}\lesssim 100\,{\rm K}$) the ion-molecule chemistry dominates, the OH formation is initiated by cosmic-ray ionization of atomic oxygen or its reaction with H$_3^+$, and the OH abundance increases with $\zeta_{_{\rm CR}}$ \citep{Meij11}. However, as discussed by \citet{Bial15}, this trend holds only up to a critical  ionization rate of $\zeta_{\rm CR, crit} \approx 10^{-14} n_3 Z'$~s$^{-1}$ (where $n_3$ is the density in units of $10^{3}\,{\rm cm}^{-3}$ and $Z'$ is the metallicity relative to Solar). For higher $\zeta_{_{\rm CR}}$ the H{\sc i}-to-H$_2$ transition occurs and the abundances of both OH and CO decrease with increasing $\zeta_{_{\rm CR}}$.

In B15, it was shown that the [CO]/[H$_2$] abundance ratio changes when varying $\zeta_{\rm CR}$ and $n_{\rm H}$ number density  (their Figures 1 and 7 respectively). The chemical analysis discussed in that work  (their \S4.1) used a gas temperature obtained from full thermal balance calculations. Comparison of our isothermal models at $T_{\rm gas}=100\,{\rm K}$ with those of \citet{Bial15} showed excellent agreement. Here, we additionally consider isothermal simulations at $T_{\rm gas}=50\,{\rm K}$ and at $20\,{\rm K}$ to explore how the [CO]/[H$_2$] ratio depends on $T_{\rm gas}$ sensitivity, a process that was left unclear in the B15 work. We complement the latter work by examining the chemical network responsible for this behaviour and what determines the [CO]/[H$_2$] ratio at different temperatures and a given $\zeta_{\rm CR}$ and $n_{\rm H}$.

\begin{figure}
\center
\includegraphics[width=0.45\textwidth]{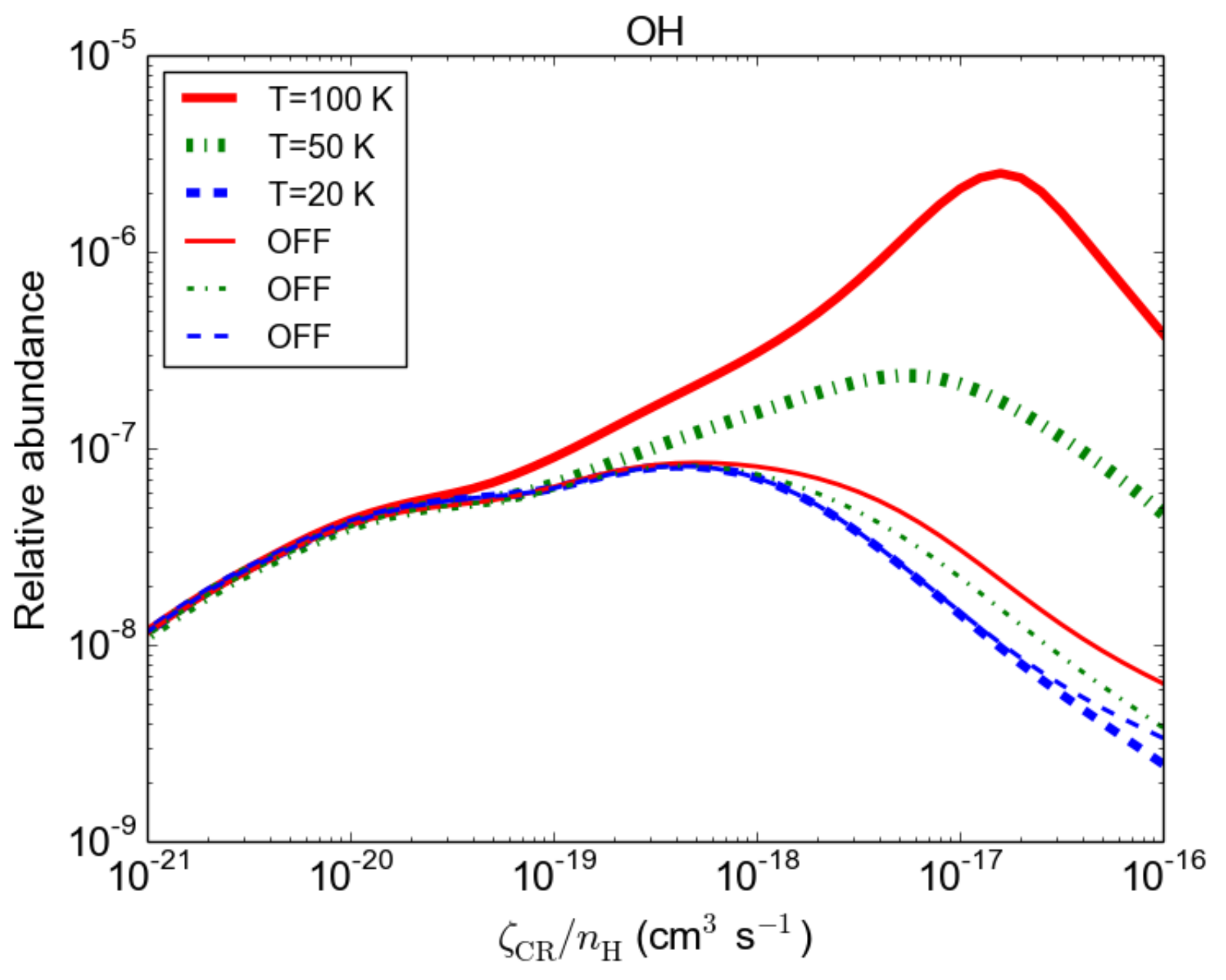}
\includegraphics[width=0.45\textwidth]{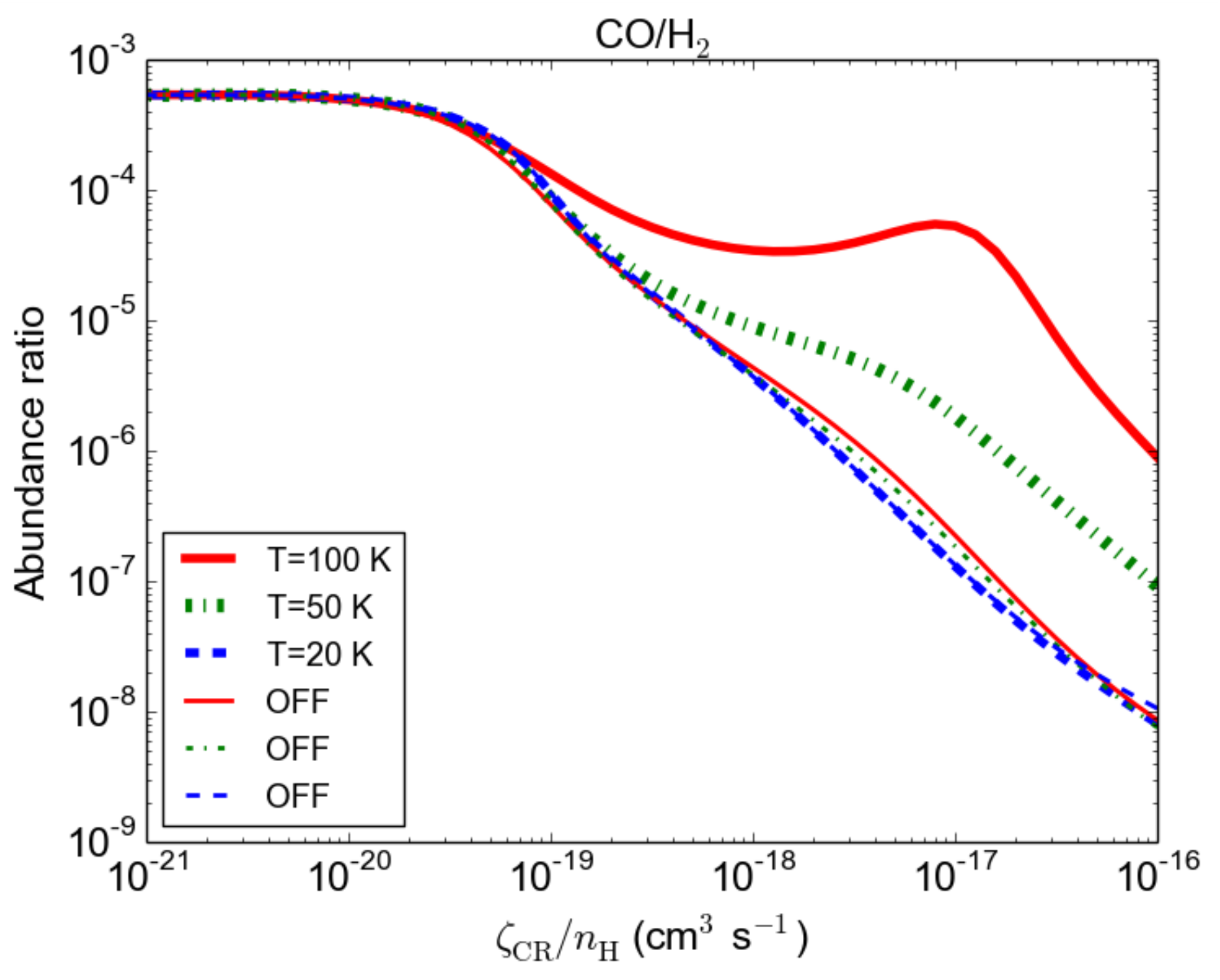}
\caption{ Isothermal test runs showing how the abundance of OH (top panel) relative to H affects the [CO]/[H$_2$] ratio (lower panel) at different temperatures. Red colour (solid lines) corresponds to $T_{\rm gas}=100\,{\rm K}$, green color (dot-dashed lines) to $T_{\rm gas}=50\,{\rm K}$, and blue colour (dashed lines) to $T_{\rm gas}=20\,{\rm K}$. In all cases, thick lines correspond to the case when Reaction \ref{ch:charge} is taken into account and thin lines when it is not (OFF). It can be seen that Reaction \ref{ch:charge} plays the dominant role controlling the [CO]/[H$_2$] ratio at different temperatures, since it triggers the formation of OH which then results in the formation of CO. By neglecting it, we do not obtain any difference in [CO]/[H$_2$] ratio as the gas temperature increases. It thus plays a key role in determining the CO abundance distribution in cosmic-ray dominated regions.}
\label{fig:chemistry}
\end{figure}

We use three different isothermal models, at $T_{\rm gas}=100\,{\rm K}$, at $50\,{\rm K}$ and at $20\,{\rm K}$ gas temperatures with $\zeta'=10^2$. Figure \ref{fig:chemistry} shows the abundances of OH (upper panel) and [CO]/[H$_2$] (lower panel) for those two different temperatures in red, green and blue colours respectively. As can be seen in the upper panel of Fig. \ref{fig:chemistry}, at $T_{\rm gas}=20\,{\rm K}$ (thick blue dashed lines), the abundance of OH slightly increases from $\zeta_{\rm CR}/n_{\rm H}\gtrsim10^{-21}\,{\rm cm}^3\,{\rm s}^{-1}$ until $\sim8\times10^{-19}\,{\rm cm}^3{\rm s}^{-1}$ at which point OH strongly decreases for an increasing $\zeta_{\rm CR}/n_{\rm H}$ ratio. As soon as $T_{\rm gas}$ is increased, the abundance of OH also increases affecting the [CO]/[H$_2$] ratio. In particular, for $T_{\rm gas}=50\,{\rm K}$ (thick dot-dashed lines) the abundance of OH keeps increasing monotonically until $\sim2\times10^{-17}\,{\rm cm}^3\,{\rm s}^{-1}$ where it peaks at an abundance of $\simeq2.5\times10^{-7}$ with respect to hydrogen. For $T_{\rm gas}=100\,{\rm K}$ (thick red solid lines), the OH abundance peaks at $\simeq3\times10^{-6}$. This trend is reflected in the [CO]/[H$_2$] abundance ratio as shown in the lower panel of this figure. In particular, for $T_{\rm gas}=20\,{\rm K}$, [CO]/[H$_2$] decreases continuously with increasing $\zeta_{\rm CR}/n_{\rm H}$. For $T_{\rm gas}=50\,{\rm K}$ and $100\,{\rm K}$, a different situation is seen: for $\zeta_{\rm CR}/n_{\rm H}\gtrsim10^{-19}\,{\rm cm}^3\,{\rm s}^{-1}$ a `turnover' appears with a local minimum at $\zeta_{\rm CR}/n_{\rm H}\sim10^{-18}\,{\rm cm}^3\,{\rm s}^{-1}$ and a local maximum at $\zeta_{\rm CR}/n_{\rm H}\sim10^{-17}\,{\rm cm}^3\,{\rm s}^{-1}$, while for higher $\zeta_{\rm CR}/n_{\rm H}$, [CO]/[H$_2$] falls.

CO forms through the OH intermediary, and OH is initiated by two important reactions; via proton transfer:
\begin{equation}
 \rm{O} + \rm{H}_3^+ \rightarrow \rm{OH}^+ + \rm{H}_2 \label{ch:proton}
\end{equation}
or via charge transfer:
\begin{equation}
\rm{O}+\rm{H}^+ \rightarrow \rm{O}^+ + \rm{H} \label{ch:charge}
\end{equation}
\begin{equation}
\rm{O}^++\rm{H}_2 \rightarrow \rm{OH}^++H
\end{equation}
\citep[see also][]{vanD86,Meij11,Bial15}. A sequence of abstraction reactions with H$_2$ followed by dissociative recombination then leads to the formation of OH \citep[see Fig. 3 of][]{Bial15}. For low gas temperatures, Reaction \ref{ch:charge} is substantially inefficient since it is endoergic by $224\,{\rm K}$ and therefore OH is mainly formed via the H$_3^+$ route (Reaction \ref{ch:proton}). CO is destroyed with He$^+$ and the reaction rate increases with increasing $\zeta_{\rm CR}/n_{\rm H}$ implying that [CO]/[H$_2$] also decreases with increasing $\zeta_{\rm CR}/n_{\rm H}$. Note that at all times as we have illustrated above, H$_2$ remains unaffected and all changes in the [CO]/[H$_2$] ratio reflect mostly the CO behaviour.

For high gas temperatures and as long as $\zeta_{\rm CR}/n_{\rm H}\lesssim10^{-19}\,{\rm cm}^{3}\,{\rm s}^{-1}$, the abundance of protons is low and therefore OH formation is dominated by Reaction \ref{ch:proton}. This makes the abundance of OH at $T_{\rm gas}=50\,{\rm K}$ and $100\,{\rm K}$ to be almost identical to that at $T_{\rm gas}=20\,{\rm K}$. In this $\zeta_{\rm CR}/n_{\rm H}$ regime, we further find that the removal of OH by C$^+$ is more efficient at low gas temperatures. Once $\zeta_{\rm CR}/n_{\rm H}\gtrsim10^{-19}\,{\rm cm}^3\,{\rm s}^{-1}$, the abundance of protons is rapidly increasing and Reaction \ref{ch:charge} becomes very efficient. This reflects the sudden increase of OH abundance as (red solid line of Fig. \ref{fig:chemistry} upper panel) and hence [CO]/[H$_2$] rises (red solid line, lower panel). Finally, for $\zeta_{\rm CR}/n_{\rm H}\sim10^{-17}\,{\rm cm}^3\,{\rm s}^{-1}$, the H{\sc i}-to-H$_2$ transition takes place and more H{\sc i} is formed. This makes the OH and consequently CO formation to become inefficient and thus both those abundances fall. 

We then perform a test to study the contribution of Reaction \ref{ch:charge} in determining the [CO]/[H$_2$] abundance ratio at different gas temperatures. To do this, we neglect this reaction by setting its rate to a negligible value and re-running the models discussed here. The resultant abundances are plotted in dashed lines in both panels of Fig. \ref{fig:chemistry}. For $T_{\rm gas}=20\,{\rm K}$ the abundances of OH and [CO]/[H$_2$] (blue dashed lines) are identical to the previous case (blue solid lines) indicating that the charge tranfer reaction is very inefficient at low temperatures. However, for higher temperatures we see that Reaction \ref{ch:charge} plays the dominant role in OH formation at high $\zeta_{\rm CR}/n_{\rm H}$ since it is primarily responsible for removing almost all protons; by neglecting it, we obtain the results of the $T_{\rm gas}=20\,{\rm K}$ test (red dashed). In turn, this is reflected in the [CO]/[H$_2$] (red dashed) as expected. This work considers Reaction \ref{ch:charge} and uses it with temperature dependency. We find that Reaction \ref{ch:charge} becomes important for gas temperatures exceeding $T_{\rm gas}\gtrsim20-30\,{\rm K}$.

Here it is important to consider that even in vigorously SF galaxies, temperatures significantly higher than $50\,{\rm K}$ may not be reached for most of their molecular gas mass. Thus the large CR-induced depressions of the average [CO]/[H$_2$] abundance ratio are expected to be maintained by the $T_{\rm gas}$-sensitive chemistry of the chemical network controlling the OH abundance. Indeed, as our Fig.~\ref{fig:cd} shows, even when $\zeta_{\rm CR}=10^3\times$Galactic (ULIRG-type of ISM), $T_{\rm gas}\la 50\,{\rm K}$. Furthermore, for metal-rich ISM environments, FUV photons cannot propagate through sufficiently high gas mass fractions to raise the average $T_{\rm gas}$ beyond that range either \citep[e.g.][]{Papa14}, while turbulent heating can only do this for minute fraction $\la 1\%$ of molecular gas mass even in the most turbulent of clouds \citep{Pan09,Pon12}. Exceptions to this will be places, such as the Galactic Center, and possibly some very extreme ULIRGs, such as Arp\,220, where $T_{\rm gas}\sim\!(50-100)\,{\rm K}$ are reached, places that either do not contain much of the total H$_2$ gas in otherwise SF-quiescent galaxies or represent SF outliers with respect to the major mode of SF in the Universe.

\section{Discussion}
\label{sec:discussion}

In this work we recover the results of B15 of a CR-induced CO destruction in H$_2$ clouds using the more realistic rendering of inhomogeneous clouds. Our three-dimensional simulations demonstrate that by increasing the cosmic-ray ionization rate, the abundance of H$_2$ molecule remains unaffected even for high cosmic-ray ionization rates of the order of $10^3$ times the mean Galactic value. On the other hand the CO abundance is sensitive to even small boosts of $\zeta_{\rm CR}$, and  is easily destroyed forming C$^+$ and consequently C (via recombination with free electrons) as long as gas temperatures $T_{\rm gas}\la 50\,{\rm K}$. Thus low-$J$ CO line emission may become very weak in such ISM environments.

Figures~2 and 3 of B15 show that the emissivities of both C lines are stronger than low-$J$ CO lines in CO-poor/H$_2$-rich regimes. This consequently yields a potential advantage of the C lines in tracing the CO-poor H$_2$ gas around the CO-rich regions of inhomogeneous H$_2$ clouds (Bisbas et al. {\it in prep.}), along with  the CO-rich H$_2$ gas. Secondary effects of a CR-induced and $n$(H$_2$)-sensitive CO destruction can make the visible H$_2$ gas distributions in SF galaxies traced by low-$J$ CO line appear clumpier than it actually is. This has been discussed by B15, but here these effects are actually computed for inhomogeneous H$_2$ clouds irradiated by elevated CR energy backgrounds (see Fig. \ref{fig:cd}). It is worth noting that the CR-induced effects studied in this work mark the warm `end' of the thermal states of potentially CO-invisible H$_2$ gas, while those affected by an enhanced Cosmic Microwave Background Radiation on low-J CO line (and dust continuum) brightness distributions at high redshifts mark the cold `end' \citep{Zhan16}. Both of these regimes may contain large amounts of molecular gas in galaxies at high ($z\ga 3$) redshifts as SF is typically a highly inefficient process, i.e. there will always be large amounts of cold non-SF H$_2$ gas and dust mass even in SF galaxies. 

\begin{figure}
\center
\includegraphics[width=0.45\textwidth]{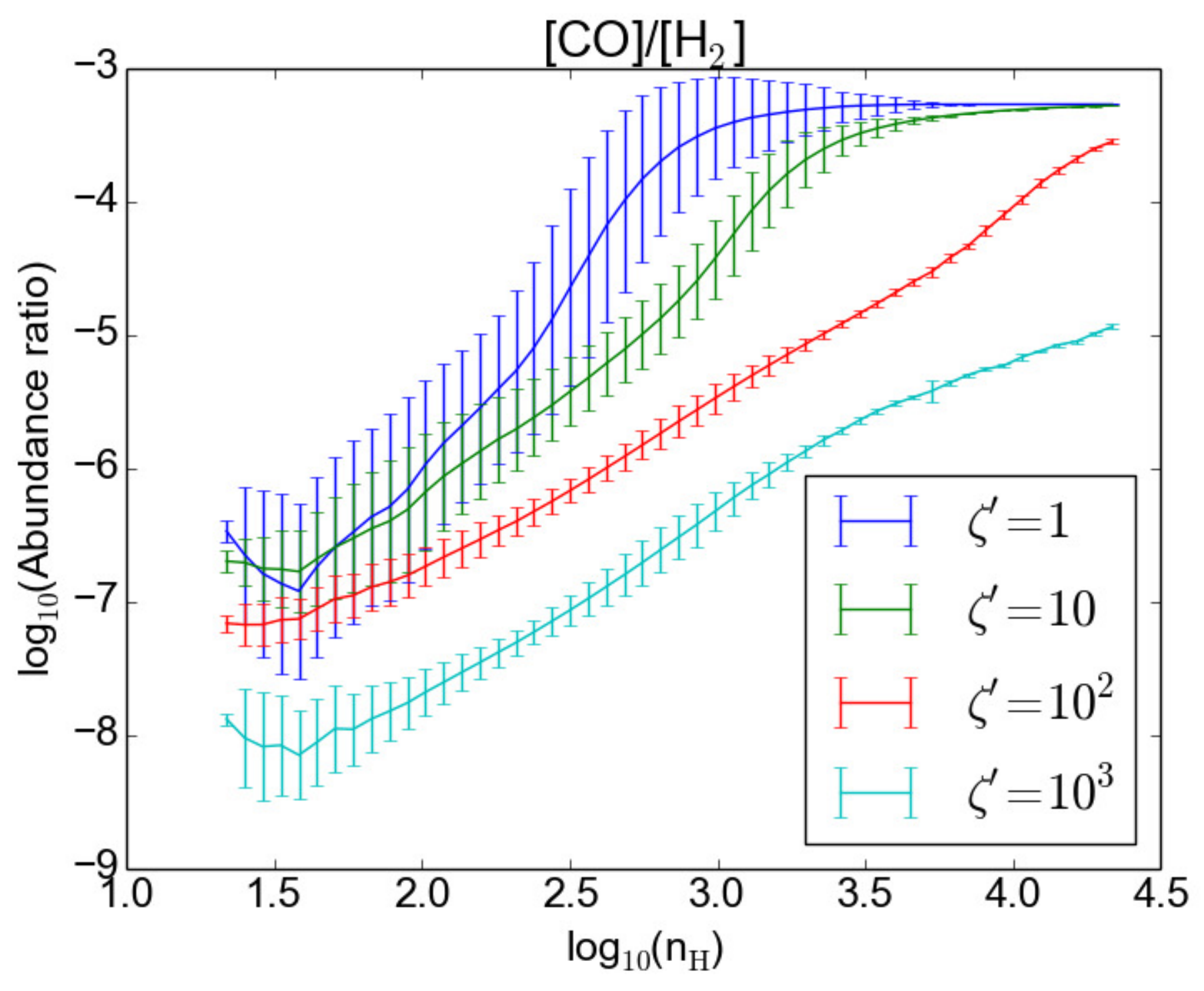}
\caption{ The [CO]/[H$_2$] abundance ratio as a function of the total H-nucleus number density, $n_{\rm H}$, for the four different $\zeta'$. For $\zeta'\lesssim10$ we do not observe any appreciable differences in the change of the ratio, whereas for $\zeta'\gtrsim10^2$ and particularly for $\zeta'\sim10^3$, almost all gas of the fractal cloud has a [CO]/[H$_2$]$<10^{-5}$ indicating that H$_2$ remains unaffected but CO has been destroyed to form mainly C and C$^+$.}
\label{fig:coh2}
\end{figure}

\subsection{Main Sequence galaxies: on the CR `firing' line}

 The destruction of the CO molecule by SF-powered CRs in H$_2$-rich galaxies is of great importance for studying star formation and its modes in the early Universe where such galaxies still strongly evolve. This is because the H$_2$ gas mass surface density and H$_2$ gas velocity dispersions are deduced from low-$J$ CO lines and are the tools to evaluate the stability of  gas-rich disks via the Q-Toomre criterion \citep{Hodg12, Dadd10}. Current theoretical views of what drives SF in strongly evolving gaseous galactic disks \citep[e.g.][]{Elme08a,Elme08b,Bour09,Bour11} depend on an accurate depiction of the H$_2$ mass surface density and velocity fields, a picture that may be incomplete because of the CR-induced CO destruction in exactly such systems.

In B15 we discussed the possibility that the average [CO]/[H$_2$] abundance may remain high in the ISM of U/LIRGs because the effect of CRs is countered by higher average molecular gas densities. However, such strong merger/starburst systems are not the main mode of SF in the early Universe. Indeed it is in massive gas-rich galaxies, evolving along a narrow region of stellar mass ($M_{*}$)-SFR plane, the so-called ``Main Sequence'' (MS) galaxies \citep{Noes07}, where $\sim\!90\%$ of the cosmic star formation takes place up to $z\sim\!3$ \citep[e.g., ][]{Bell05,Elba07,Rodi11}.  It is these systems, with ${\rm SFR}\sim\!(20-300)\,{\rm M}_{\odot}\,{\rm yr}^{-1}$ \citep[e.g][]{Genz12} and seemingly ordinary GMCs \citep{Carl16}, that are expected to be the most affected by a CR-induced destruction of CO. This is apparent from Fig.~1 of B15 (where BzK galaxies are MS systems) as well as Fig.~\ref{fig:cd} of this work from where it can be seen that for ${\rm SFR}\!\sim\!(10-100)\,{\rm M}_{\odot}\,{\rm yr}^{-1}$, which may correspond to $\zeta'\sim10-10^2$, the CO `marking' of a typical molecular cloud  is significantly reduced.

The metallicity-insensitive CR-induced destruction of CO in MS galaxies can only compound the difficulties already posed by the lower metallicities prevaling in some of these systems \citep[][and references therein]{Genz12}, making any CO-deduced H$_2$ gas mass distributions, their scale-lengths, their SFR-controlled gas depletion timescales, dynamical masses and Q-Toomre stability criteria, provisional. In this context it is important to remember that even the well-known effects of strong FUV/low-Z on the [CO]/[H$_2$] abundance ratio can render entire clouds CO-free \citep{Bola99,Pak98} boosting their C (and C$^+$) content. In the $Z-\chi_0$/CR domain where a phase transition to very CO-poor H$_2$ gas phases happens, it will be highly non-linear, making a practical calibration of the so-called $X_{\rm CO}$ factor in MS galaxies \citep{Dadd10,Genz12,Carl16} and local spiral LIRGs \citep{Papa12a} challenging, even for their CO-marked H$_2$ gas distributions.

Indeed as we have already discussed in \S\ref{ssec:3dpdr}, the ability of CO to survive only in the densest regions of CR-irradiated clouds (Fig.~\ref{fig:cd}) can yield a misleading picture about the actual $\rm H_2$ gas distribution and its thermal and dynamical state. Moreover, a nearly-Galactic $X_{\rm CO}$ factor may still be obtained if such CO-marked sub-regions of the underlying H$_2$ distribution are used for its calibration, even as they are no longer representative of the actual H$_2$ clouds. This can be shown if we consider the dense H$_2$ gas regions where CO survives embedded in columns of CO-free $\rm H_2$ gas. The latter will exert a non-thermal pressure of 
\begin{equation}
P_{\rm e} \approx \frac{\pi}{2} G \Sigma ({\rm H}_2)\left[\Sigma ({\rm H}_2)+\left(\frac{\sigma_{g}(V)}{\sigma_{*}(V)}\right)\Sigma_{*}\right],
\end{equation}
where we assumed that the CO-rich gas regions lie mid-plane in a rotating $\rm H_2$-rich disk, with stars mixed in, at a surface mass density of $\rm \Sigma_{*}$ and vertical velocity dispersion of $\rm \sigma_{*}(V)$ (with $\Sigma ({\rm H}_2)$ and $\sigma_{g}(V)$ the corresponding quantities for the $\rm H_2$ gas). For Milky Way, $P_{\rm e}/k_{\rm B}\sim 1.4\times 10^{4}\,{\rm cm}^{-3}\,{\rm K}$ is the average non-thermal pressure on the boundaries of molecular clouds \citep{Elme89}. This, crucially, determines the normalization of the so-called linewidth-size relation for a molecular cloud of radius $R$:
\begin{equation}
\sigma (R) = \sigma _0\left(\frac{P_{\rm e}/k_{\rm B}}{10^4\,{\rm K}\,{\rm cm}^{-3}}\right)^{1/4}\left(\frac{R}{\rm pc}\right)^{1/2}\,
\end{equation}
\citep{Chie87,Elme89}. Should a CO-invisible H$_2$ gas mass lie `on-top' of CO-rich (and thus observable) cloud regions in the mid-plane of a SF disk, it would exert an `overpressure' on the CO-rich ones. This  would appear as a deviation from a Galactic line-width size relation and  CO clouds that would seem to be out of virial equillibrium, lowering their corresponding $X_{\rm CO}$ factor \citep{Down98,Papa12a}. Nevertheless the very  weak dependance of the line-width size relation on $P_{\rm e}$ allows large amounts of CO-invisible gas to exist without easily discernible observational effects. For a purely H$_2$  gas disk ($\Sigma_{*}=0$) (assumed here for simplicity), a $\Sigma_{g}{\rm (CO-invisible)}=5\times\Sigma_{g}{\rm(CO-visible)}$ larger H$_2$ gas surface density would raise $P_{\rm e}$ by a factor of 25, but the corresponding $\rm \sigma (R)$ of CO-rich clouds embedded inside such overlying columns of CO-invisible gas only by a factor of $\sim\!2.2$. The latter is within the observational uncertainty of the $\sigma(R)$ relation in the Galaxy \citep[e.g.][]{Heye04} and thus any $X_{\rm CO}$ calibration of such overpressured CO-rich clouds would still give a value consistent with a Galactic one within the uncertainties.

\subsection{CO chemistry in SF galaxies: towards a dynamical framework}

 The physics and chemistry of the CR-induced destruction of CO can now be readily used in  galaxy-sized/cosmological evolution models. Doing so will:  a) shed light on what happens in the context of galaxy evolution models as the $\chi_0$/CR `boundary' conditions of molecular clouds evolve, and b) help the interface of such models with actual observables (e.g. low-$J$ CO and C images of galaxies with ALMA/JVLA). Past work has already incorporated, in a sub-grid fashion, the effects of FUV destruction of CO in H$_2$ clouds inside galaxies \citep[]{Pelu09,Olse15,Rich16}.  CR-driven effects will be even easier to implement in such models insofar as full transparency of H$_2$ clouds to CRs and $U_{\rm CR}\propto\rho_{\rm SFR}$ are assumed. Moreover, C lines as H$_2$ gas mass tracers in galaxies at high redshifts have already been discussed in a cosmological context \citep{Toma14}. Once CR effects are taken into account in galaxy-scale models, one can then: a) evaluate the best method(s) in obtaining the H$_2$ mass distributions and velocity fields in SF galaxies evolving across cosmic epoch, and b) ascertain whether current theoretical views about the role of unstable  giant H$_2$ clumps in driving the SF of gas-rich early galaxies \citep{Bour14} still hold (e.g. a C-imaged $\Sigma ({\rm H}_2)$ distribution may be a smoother one than a CO-imaged one in a CR-irradiated gaseous disk, impacting also the deduced gas velocity fields from these lines).

Such models can also shed light on another very important caveat discussed by B15, namely the role of turbulence. Observations of the so called (U)LIRGs, extreme merger/starburst systems, indicate that regions with high SFR density (and thus $\zeta_{\rm CR}$) are also regions of strong turbulence of H$_2$ clouds, and thus of high $M(n_{\rm H}>10^4\,{\rm cm}^{-3})/M({\rm H}_2)$ mass fractions per GMC. With CO remaining abundant in high-density gas ($n_{\rm H}>10^4\,{\rm cm}^{-3}$), even when average $\zeta_{\rm CR}$ are high, this can diminish and even counteract  the effects of CR-induced CO destruction in such environments as now most of the H$_2$ gas no longer resides in the low-density regime ($\sim10^2-10^3\,{\rm cm}^{-3}$) as in the MW (densities where CO would be CR-destroyed very effectively) but at high densities. In this regard, the [CO]/[H$_2$] abundance ratio obtained from our simulations is of particular interest. In Fig.~\ref{fig:coh2} we plot this ratio versus the total H-nucleus number density for all four different $\zeta'$ examined. For $\zeta'\gtrsim10^2$ most of the molecular cloud gas has [CO]/[H$_2$]$<10^{-5}$, making it very CO-poor. For higher densities and according to B15, this ratio would exceed $10^{-5}$ at $\zeta'=10^3$ only for $n_{\rm H}>10^4\,{\rm cm}^{-3}$ assuming no significant freeze-out. 

Numerical simulations of individual turbulent H$_2$ clouds study the effects of constant, pre-set FUV radiation fields and CR energy densities on the CO and C distributions and the corresponding line emission \citep[e.g.][as well as the present work]{Glov16}. Such models, while useful in finding trends of CO and C line emission as H$_2$ gas mass tracers in GMCs, {\it they cannot address the issue of what happens when such clouds are immersed in actual galaxies,} where the FUV and CR energy densities around these clouds vary strongly on timescales equal or shorter than internal cloud chemical and dynamical timescales \citep{Pelu06}. This is because in individual cloud simulations the `boundary' conditions of FUV radiation, $\zeta_{\rm CR}$, and turbulent energy injection  are not tracked. Galaxy-sized models \citep[e.g.][]{Smit14} that include H$_2$ clouds, along with the appropriate physics and chemistry behind the FUV/CR `drivers', modelled {\it in tandem} with the evolving conditions of a SF galaxy are thus invaluable in examining whether in high SFR-density environments, H$_2$ gas remains mostly CO-rich or not \citep[see][for an early such treatment where CO and H$_2$ are treated separately and FUV radiation onto clouds within galaxies is tracked]{Pelu09}.

Regardless of any future theoretical `verdict' on  whether CO-invisible molecular gas can exist in large quantities in SF galaxies in the early Universe during periods of high SFR densities, observations of low-$J$ CO \emph{and} C lines in such systems are indispensable. Here we re-iterate that the C column density retains its robustness in tracing H$_2$ column near-proportionally for $\zeta_{\rm CR}\sim10^{-15}\,{\rm s}^{-1}$ (see Fig. \ref{fig:cd}). In particular we find that for such CR-ionization rates, $N({\rm C}${\sc i}$)\simeq4\times10^{-4}\,N({\rm H}_2)$ (for our adopted carbon elemental abundance). This relation depends weakly on $\zeta_{\rm CR}$ provided that $\zeta_{\rm CR}\gtrsim10^{-15}\,{\rm s}^{-1}$, contrary to the corresponding one for CO. Moreover, even for MW-level of $\zeta_{\rm CR}$ values, the $W_{\rm CI,1-0}$ per beam will remain larger than that of CO $J=1-0$ or $J=2-1$ (the two CO transitions used to trace the bulk of H$_2$ mass in galaxies) as long as the same beam is used to image the H$_2$ gas in all lines\footnote{The lowest observed brightness temperature ratio of $W_{\rm CI,1-0}$/$W_{\rm CO(1-0)}$ in Milky Way is $\sim0.1$ \citep{Papa04}. If we were to observe $W_{\rm CI,1-0}$ and $W_{\rm CO(1-0)}$ at the same resolution, then in the Rayleigh-Jeans regime the flux per beam boost would be $(492\,{\rm GHz}/115\,{\rm GHz})^2\times0.1\sim1.83$. This can lead to signal-to-noise advantages for the C line obsrevations depending on the redshift of the object \citep[see][]{Papa04}. If $\zeta'$ is increased, C lines are becoming brighter still, increasing this kind of advantage.}. This, along with the possibility that C line imaging of SF disks at high-$z$ finds a different (smoother and/or possibly more extended) H$_2$ gas distribution, because of large quantities of CO-poor H$_2$, argues strongly for sensitive C line imaging of gas-rich SF galaxies (Bisbas et al. {\it in preparation)}. 

In this work we also identified the exact chemistry behind the large gas temperature sensitivity on the CO formation in H$_2$ clouds (see \S\ref{sec:oh}), an issue initially discussed by B15. The gas temperature sensitivity of our results elevates the importance of reliable computation of the average thermal state of H$_2$ gas in FUV/CR-intensive environments found within SF galaxies. As demonstrated in Fig.~\ref{fig:heat}, CRs can provide an important heating source throughout the GMC and particularly in regions with $n_{\rm H}>10^3\,{\rm cm}^{-3}$. This in turn can increase the gas temperature deep in the cloud (where the FUV has been severely attenuated) to values of the order of $\sim50\,{\rm K}$ \citep[][]{Meij11,Papa11}. Such gas temperatures are still low for the CO formation to occur via the O+H$^+$ charge transfer. In low metallicity galaxies, a higher gas temperature may be expected as cooling efficiency and shielding is less than in solar metallicity ones, perhaps moderating the CR-induced destruction of CO in high $\zeta'$ environments, as discussed in \S\ref{sec:oh}.  It is thus necessary that the CR effects  are  studied together with those driven by  lower metallicities  in order  to discern their combined impact on  the average [CO]/[H$_2$] abundance in metal-poor star forming galaxies.

Even though we used  standard cooling/heating mechanisms of PDR/CRDR physics (see \S\ref{ssec:heatcool}), turbulence will also heat the molecular gas \citep[e.g.][]{Pan09}, and do so in a volumetric manner just like the CRs. Turbulent heating has even been argued as a dominant heating mechanism of galaxy-sized H$_2$ gas reservoirs in some extreme SF galaxies \citep{Papa12b}, even as numerical simulations of individual molecular clouds show that turbulent heating typically affects only $\la 1\%$ of their mass \citep{Pan09,Pon12}. The so-called `Brick' cloud is a well studied object close to the Galactic Centre. Simulations performed by \citet{Clar13} have reproduced its observed gas and dust temperatures when their modeled cloud interacts with a FUV of strength $\chi/\chi_0\sim10^3$ and a $\zeta_{\rm CR}\sim10^{-14}\,{\rm s}^{-1}$. Early suggestions by \citet{Ao13} proposed that the gas heating of the Central Molecular Zone (CMZ) at the Galactic Centre is primarily dominated by cosmic-rays and/or turbulence. Recent observations by \citet{Gins16}, however, show that the dominant heating mechanism of the particular `Brick' cloud is turbulence which, in association with an LVG analysis, give an upper limit of $\zeta_{\rm CR}<10^{-14}\,{\rm s}^{-1}$. However, in places of the CMZ cosmic-rays may still be a very important heating source particular in less turbulent sub-regions \citep{Gins16}.

Our simple treatment of turbulent heating \citep{Blac87,Rodr01,Bisb12} leaves unaswered the question of how much it can influence the average thermal states of the typically very turbulent H$_2$ gas in extreme starbursts with high SFR densities. Numerical simulations of individual H$_2$ gas clouds, at Mach numbers appropriate for the ISM  of galaxies with very high SFR densities (${\cal M}\sim3-10$ times that of ordinary spirals), that include turbulent heating along with the chemistry and physics of CR-induced CO destruction are necessary for answering this question. If strong turbulence can elevate the {\it average} H$_2$ gas temperatures and densities of galaxies with high  SFR densities (typically merger/starbursts), it may still keep the [CO]/[H$_2$] abundance ratio high and the H$_2$ gas traceable via the traditional methods based on CO (see B15 for the relevant discussion).

\section{Conclusions}
\label{sec:conclusions}%

In this paper, continuing the study of \citet{Bisb15}, we present results from a suite of three-dimensional astrochemical simulations of inhomogeneous molecular clouds, rendered as a fractal, and embedded in different cosmic-ray ionization rates spanning three orders of magnitude ($\zeta_{\rm CR}=10^{-17}-10^{-14}\,{\rm s}^{-1}$) along with a constant isotropic FUV radiation field ($\chi/\chi_0=1$). Our study therefore focuses only on the effect of  high cosmic-ray ionization rates expected in SF galaxies in the Universe, and how it affects the abundances of CO, C, C$^+$, H{\sc i} and H$_2$. We used the {\sc 3d-pdr} \citep{Bisb12} code to perform full thermal balance and chemistry calculations. Our results can be summarized as follows:

\begin{enumerate}

\item The column density and  total H$_2$ mass of a typical inhomogeneous GMC remains nearly constant for increasing $\zeta'$, with the total mass of H$_2$ decreasing  by $\lesssim 10\%$ for $\zeta_{\rm CR}\sim10^{-14}\,{\rm s}^{-1}$ ($\rm \sim 10^{3}\times Galactic$). On the other hand  a significant reduction of the [CO]/[H$_2$] abundance ratio sets in throughout the cloud, even when $\zeta_{\rm CR}\sim10^{-16}\,{\rm s}^{-1}$ ($\rm \sim 10\times Galactic$), a value expected for the ISM of many star-forming galaxies in the Universe.

\item When the average $\zeta_{\rm CR}$ increases further, up to $\sim 10^{-15}-10^{-14}\,{\rm s}^{-1}$ the CO molecule is destroyed so thoroughly that only the densest regions of the GMC remain CO-rich. The abundances of C and C$^+$ on the other hand increase, with the latter becoming particularly abundant for $\zeta_{\rm CR}\sim 10^{-14}\,{\rm s}^{-1}$. Atomic carbon is the species that proves to be the most abundant, `marking' most of the H$_2$ mass of the cloud over a wide range of $\zeta_{\rm CR}$ values. Using only CO rotational transitions to discern the average state and mass of such CR-irradiated GMCs will only recover their highest density peaks ($n_{\rm H}\gtrsim10^3\,{\rm cm}^{-3}$), make the clouds appear clumpier than they truly are, and convey biased information on the molecular gas velocity fields. 
 
\item We expect significant effects of CR-induced destruction of CO to occur in the so-called Main Sequence Galaxies, the systems where most of the cosmic history of star formation unfolds. This is a result of their high SF rates (implying high CR rates) and seemingly Galactic-type molecular clouds. The  widespread CR destruction of CO expected in such systems will make the calibration of their $X_{\rm CO}$ factor challenging, even for their CO-bright gas.

\item Our computations recover gas temperatures of $T_{\rm gas}\sim10\,{\rm K}$ for the CR-irradiated and FUV-shielded dense regions inside those GMCs. This is indeed typical for such regions in the Galaxy and it remains robust over $\zeta_{\rm CR}\lesssim10^{-16}\,{\rm s}^{-1}$. This is of particular importance if the initial conditions of SF, and the stellar initial mass function (IMF) mass scale (i.e. the IMF `knee') are indeed set within such regions. Nevertheless, once $\zeta_{\rm CR}\sim10^{-15}-10^{-14}\,{\rm s}^{-1}$, the temperature of such regions rises up to $T_{\rm gas}\sim30-50\,{\rm K}$, and the initial conditions of star formation in such galaxies are bound to change.

\item The main heating mechanisms in cosmic-ray dominated regions apart from CRs, are the chemical mechanism (due to the large amounts of ions expected in CRDRs) and the H$_2$ formation mechanism. Cooling on the other hand, is mainly due to C$^+$ and O with the contribution of CO cooling nearly negligible, as its abundance is at least two orders of magnitude lower than in normal Galactic conditions.

\item We find the CR-regulated [CO]/[H$_2$] abundance ratio to be sensitive to the temperature of the gas once $\rm T_{gas}>50\,K$. A significant production of the OH molecule, acting as an intermediary, is the $T_{\rm gas}$-sensitive part of the chemical network that determines the [CO]/[H$_2$] ratio. For warm gas at $T_{\rm gas}=100\,{\rm K}$ abundant OH can keep the molecular gas CO-rich (i.e. [CO]/[H$_2$]$\sim 10^{-4}$), even in high CR energy environments. The severe CR-induced destruction of CO sets in for $\rm T_{gas}\la 50 K\,$, which our thermochemical calculations indicate as containing the bulk of H$_2$ mass in our inhomogeneous cloud models, and indeed the bulk of molecular gas in SF galaxies, except perhaps in the most extreme merger/starbursts.

\item Our simple treatment of turbulent heating, and the fact that GMCs in the very high SFR density environments of merger/starburst galaxies are much more turbulent and thus denser, necessitate careful considerations of turbulent heating and a dynamic rendering of density inhomogeneities in order to explore our findings in a fully realistic setting. 

\item Finally, the chemistry and thermal-balance calculations behind the CR-controlled [CO]/[H$_2$], [C]/[H$_2$], and [C$^+$]/[H$_2$] abundance ratios inside inhomogeneous H$_2$ clouds can be used in a sub-grid fashion as elements of galaxy-sized numerical simulations of evolving galaxies. This is perhaps a vital ingredient of any realistic galaxy evolution model across cosmic epoch, given the elevated SFR densities --and thus CR energy densities-- typically observed in galaxies in the distant Universe.

\end{enumerate}

As a final conclusion we  mention that because of the  strong effects of CRs on the CO abundance combined with the  effects of high FUV and/or low metallicity environments in further reducing its abundance, and the  impracticallity of C$^{+}$ imaging in SF galaxies except for the highest redshift objects ($z\ga 4$), a concerted effort must be mounted by the extragalactic community towards C line imaging of H$_2$ gas in the Universe as a viable alternative.

\section*{Acknowledgements}
The authors thank an anonymous referee for reviewing the manuscript and whose comments have impoved the clarity of this work. We thank Andreas Schruba, Andrew Strong, Rob Ivison, Nick Indriolo, Steffi Walch and Paola Caselli for the useful discussions. This work is supported by a Royal Netherlands Academy of Arts and Sciences (KNAW) professor prize, and by the Netherlands Research School for Astronomy (NOVA). The work of PPP was funded by an Ernest Rutherford Fellowship. SB acknowledges support from the DFG via German-Israel Project Cooperation grant STE1869/2-1 GE625/17-1. LSz acknowledges support from the A-ERC grant 108477 PALs. ZYZ acknowledges support from ERC in the form of the Advanced Investigator Programme, 321302 COSMICISM. 

\appendix

\section{A. Chemical network and initial elemental abundances}
\label{app:chem}

\begin{figure}
\center
\includegraphics[width=0.33\textwidth]{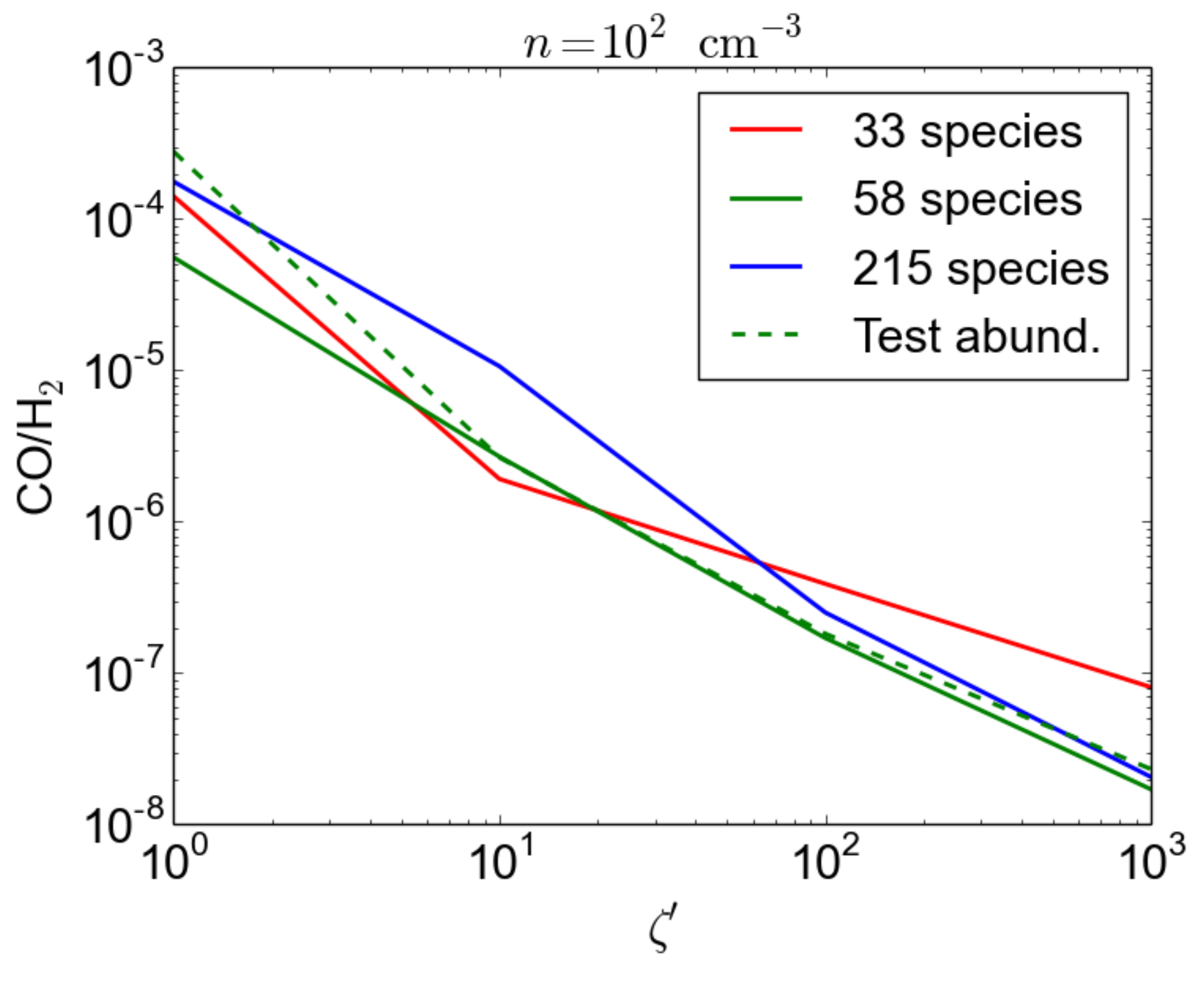}
\includegraphics[width=0.33\textwidth]{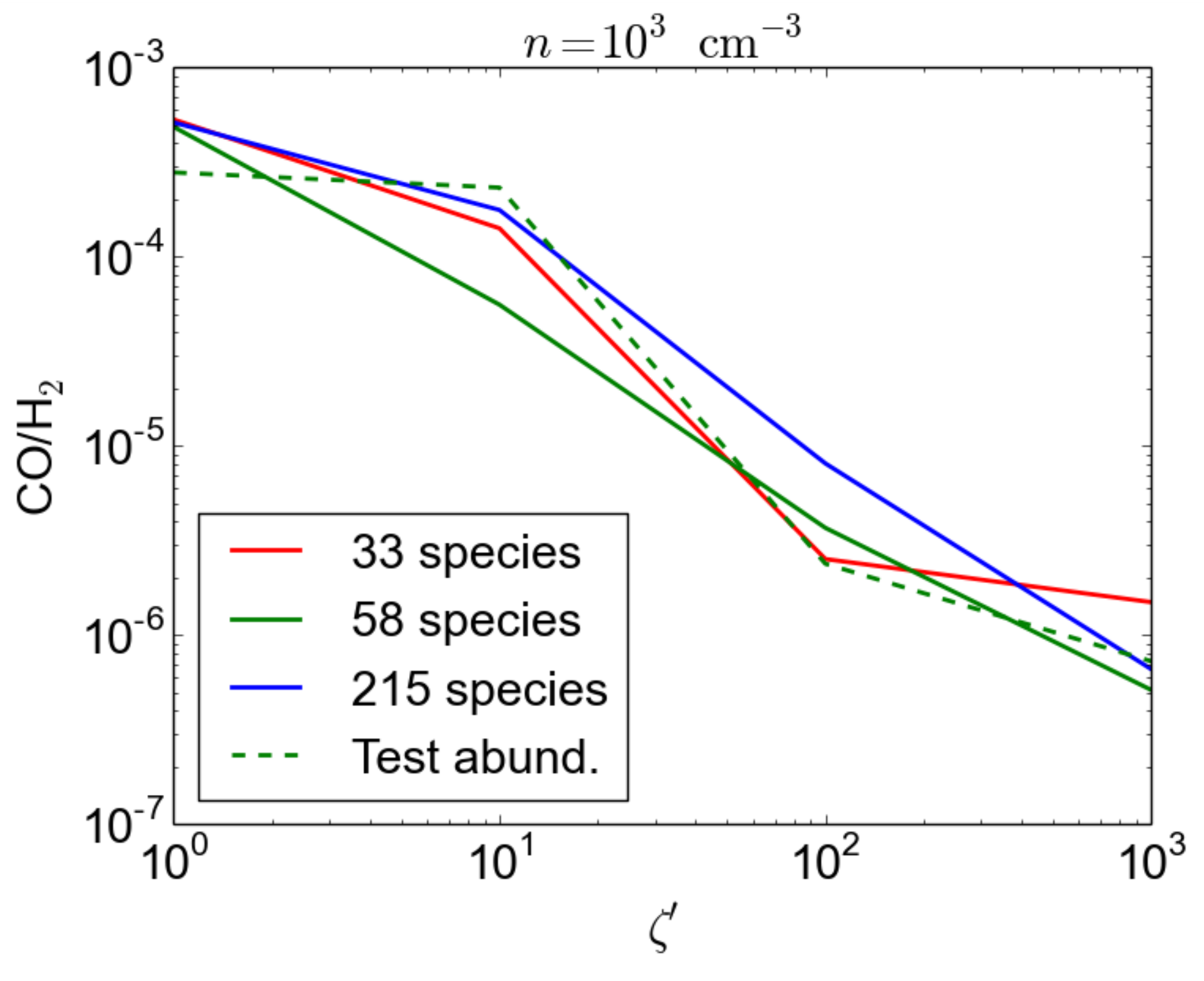}
\includegraphics[width=0.33\textwidth]{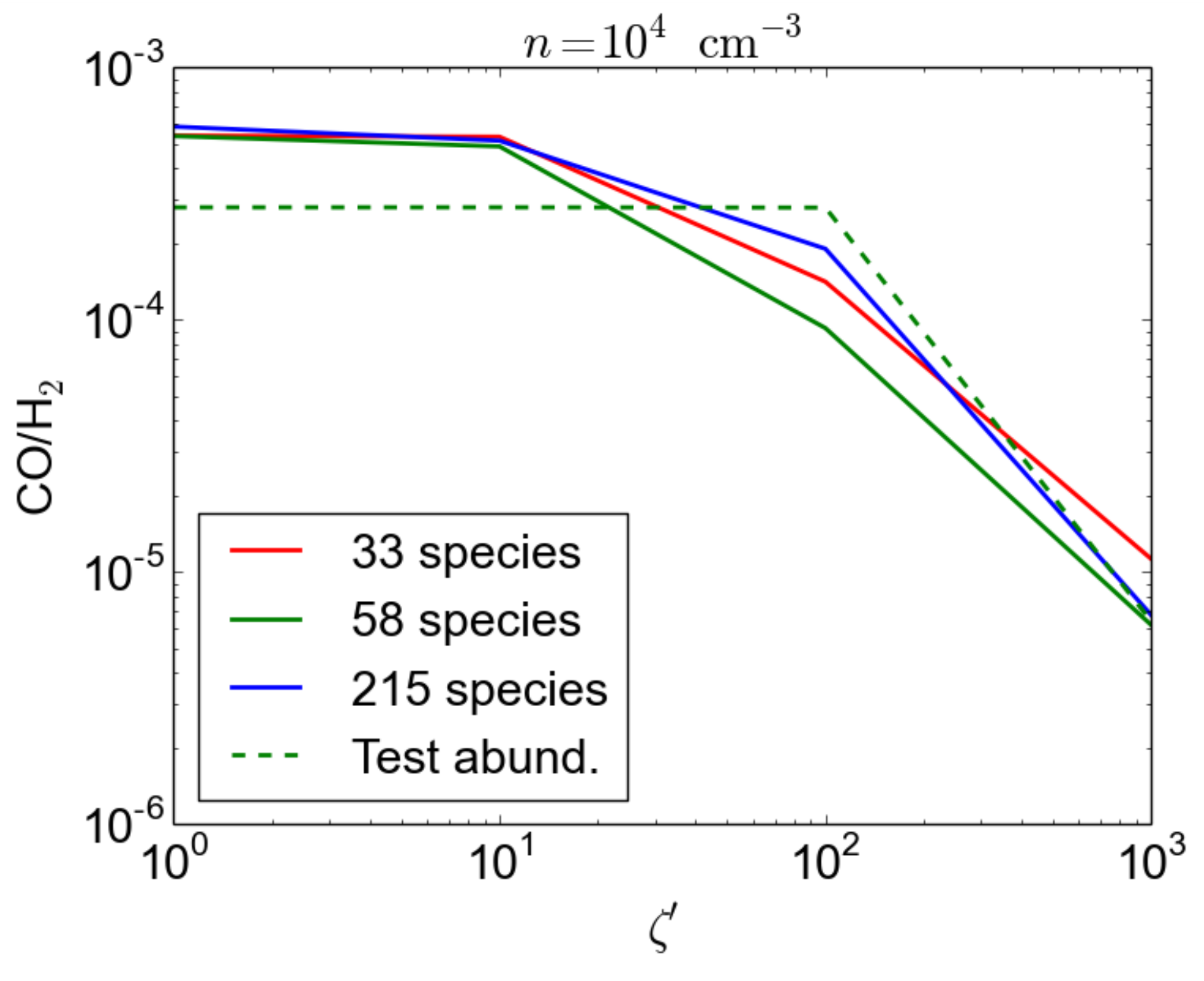}
\label{appfig}
\caption{ Dependency of the [CO]/[H$_2$] abundance ratio on $\zeta'$ for $n_{\rm H}=10^2\,{\rm cm}^{-3}$ (left), $10^3\,{\rm cm}^{-3}$ (middle), and $10^4\,{\rm cm}^{-3}$ (right) using a 33-species (red solid line), 58-species (green solid line) and 215-species (blue solid line) chemical network with initial abundances as discussed in \citet{Aspl09}. The green dashed line (``Test abund.'') corresponds to the 58-species network with initial elemental abundances as discussed in Appendix \ref{app:chem}. We find that the general trend of [CO]/[H$_2$] decrease by the increase of cosmic-ray ionization rate remains robust supporting the validity of our results.}
\end{figure}

We explore the dependence of our results on the choice of chemical network and the choice of initial elemental abundances. To do this, we perform a suite of 0D calculations where we switch off the UV radiation field. We use densities of $n_{\rm H}=10^{2-4}\,{\rm cm}^{-3}$ interacting with $\zeta'=10^{0-3}$ cosmic-ray ionization rates (see Eqn.~\ref{eqn:zeta'}). We consider two subsets and the full UMIST 2012 \citep{McEl13} consisting of 33 species (4 elements: H, He, C, O), 58 species (2 additional elements: Mg, S), and 215 species (4 additional elements: Na, Fe, Si, N) respectively. In addition to Table~\ref{tab:abun}, the initial abundances of the last four elements used are Na=1.738$\times$10$^{-6}$, Fe=3.162$\times$10$^{-5}$, Si=3.236$\times$10$^{-5}$ and N=6.76$\times$10$^{-5}$ \citep{Aspl09}. The results of the above tests are shown in red solid (33 species), green solid (58 species) and blue solid (215 species) lines in Fig.~\ref{appfig}1. 

We further perform additional simulations using the subset of 58 species only and in which we change the initial values of elemental abundances to those that have been measured by optical/UV absorption lines in diffuse clouds with densities similar to those of our fractal GMCs. We use C=1.4$\times$10$^{-4}$ \citep{Card96}, O=2.8$\times$10$^{-4}$ \citep{Cart04} and Mg=7$\times10^{-9}$ while keeping S as shown in Table~\ref{tab:abun} as it is observed to remain largely undepleted. The reduction of the Mg abundance by $\sim4$ orders of magnitude compared to the value shown in Table~\ref{tab:abun} is motivated by the fact that such high abundances of Mg may act as a non-negligible source of electrons which can in turn affect the [CO]/[H$_2$] ratio. The results of this test are shown in Fig.~\ref{appfig}1 as green dashed lines. These abundances correspond to environments with metallicities $Z\simeq Z_{\odot}$. We note that thoughout this work we have assumed solar metallicity at all times and we do not explore the effect of CR-induced CO destruction in sub-solar and super-solar environments. For all reasonable assumptions, C/O$\sim0.5$ which is consistent with diffuse ISM observations, and always $<1$.

Overall, from the above suite of tests we find that the general trend of the [CO]/[H$_2$] abundance ratio decrease by increasing the cosmic-ray ionization rate remains robust. We therefore demonstrate that the validity of findings presented in this work and in particular the column density maps shown in Fig. \ref{fig:cd} do not strongly depend on the complexity of the chemical network used or the choice of initial elemental abundances adopted.

\section{B. Mapping SPH to grid}
\label{app:sph2grid}

We convert the properties of the cloud (number density distribution, gas temperatures, etc.) from SPH to uniform grid in order to produce the column density plots of Fig. \ref{fig:cd}. Each SPH particle, $p$, comprising the cloud has a smoothing length $h_p$ and carries the corresponding PDR information from the {\sc 3d-pdr} calculations. In order to weight an SPH quantity, $A_p$, in the centroid, $q$, of a given cell of the uniform grid we use the equation
\begin{eqnarray}
A_q=\sum_{p=1}^Nn_pA_pW(\left|{\bf r_q-r_p}\right|,h_p),
\end{eqnarray}
where $N=50$ is the number of the closest neighbouring SPH particles to the centroid of the cell and $W$ is the \citet{Mona85} softening kernel
\begin{eqnarray}
W(\ell,p)=\frac{1}{\pi h_p^3}\begin{cases}
   1-\frac{3}{2}\ell^2+\frac{3}{4}\ell^3, & {\rm if}\,0\le\ell<1\\
   \frac{1}{4}(2-\ell)^3, & {\rm if}\,1\le\ell\le2\\
   0, & {\rm if}\,\ell>0,\\
\end{cases}
\end{eqnarray}
where $\ell=\left|{\bf r_q-r_p}\right|/h_p$. The number of SPH particles in each grid cell varies from few tens (highest density regions) to none (outside the cloud). Similar techniques have been discussed by \citet{Pric07}.

\end{document}